\documentclass[manuscript=article,journal=jpcafh]{achemso}
\usepackage[dvipsnames]{xcolor}
\usepackage{subcaption}
\usepackage{amssymb}
\usepackage{amsmath}
\usepackage{braket}
\usepackage{multirow}
\usepackage{adjustbox}
\usepackage{array}
\usepackage{bm}
\newcolumntype{L}[1]{>{\raggedright\let\newline\\\arraybackslash\hspace{0pt}}m{#1}}
\newcolumntype{C}[1]{>{\centering\let\newline\\\arraybackslash\hspace{0pt}}m{#1}}
\newcolumntype{R}[1]{>{\raggedleft\let\newline\\\arraybackslash\hspace{0pt}}m{#1}}
\usepackage{pdfpages}
\usepackage[version=3]{mhchem}
\usepackage{xr}

\newcommand{\rev}[1]{{\color{black}#1}}  

\makeatletter
\newcommand*{\addFileDependency}[1]{
  \typeout{(#1)}
  \@addtofilelist{#1}
  \IfFileExists{#1}{}{\typeout{No file #1.}}
}
\makeatother

\newcommand*{\myexternaldocument}[1]{%
    \externaldocument{#1}%
    \addFileDependency{#1.tex}%
    \addFileDependency{#1.aux}%
}

\myexternaldocument{SI}

\title{A Universal Graph Deep Learning Interatomic Potential for the Periodic Table}

\author{Chi Chen}
\affiliation{Department of NanoEngineering, University of California San Diego, CA, USA
}
\email{chenc273@outlook.com}

\author{Shyue Ping Ong}%
\affiliation{Department of NanoEngineering, University of California San Diego, CA, USA
}
\email{ongsp@eng.ucsd.edu}

\date{\today}

\begin{document}

\maketitle

\begin{abstract}
\textbf{Interatomic potentials (IAPs), which describe the potential energy surface of atoms, are a fundamental input for atomistic simulations. However, existing IAPs are either fitted to narrow chemistries or too inaccurate for general applications. Here, we report a universal IAP for materials based on graph neural networks with three-body interactions (M3GNet). The M3GNet IAP was trained on the massive database of structural relaxations performed by the Materials Project over the past 10 years and has broad applications in structural relaxation, dynamic simulations and property prediction of materials across diverse chemical spaces. About 1.8 million materials were identified from a screening of 31 million hypothetical crystal structures to be potentially stable against existing Materials Project crystals based on M3GNet energies. Of the top 2000 materials with the lowest energies above hull, 1578 were verified to be stable using DFT calculations. These results demonstrate a machine learning-accelerated pathway to the discovery of synthesizable materials with exceptional properties.}
\end{abstract}

Atomistic simulations are the bedrock of \textit{in silico} materials design. The first step in most computational studies of materials is obtaining an equilibrium structure, which involves navigating the potential energy surface (PES) across all independent lattice and atomic degrees of freedom in search of a minimum. Atomistic simulations are also used to probe the dynamical evolution of materials systems and to obtain thermodynamic averages and kinetic properties (e.g., diffusion constants). While electronic structure methods such as density functional theory (DFT) provide the most accurate description of the PES, they are computationally expensive and scale poorly with system size. 

For large-scale materials studies, efficient, linear-scaling interatomic potentials (IAPs) that describe the PES in terms of many-body interactions between atoms are often necessary. However, most IAPs today are custom-fitted for a very narrow range of chemistries, often for a single element or up to no more than 4-5 elements. The most popular ``general purpose'' IAPs are the AMBER family of force fields\cite{weinerAMBERAssistedModel1981,caseAmberBiomolecularSimulation2005} and the Universal Force Field (UFF)\cite{rappeUFFFullPeriodic1992}. However, both were formulated primarily for molecular/organic systems and have limited support and accuracy in modeling crystal structures. More recently, machine learning (ML) of the PES has emerged as a particularly promising approach to IAP development.\cite{behlerGeneralizedNeuralNetworkRepresentation2007,bartokGaussianApproximationPotentials2010,thompsonSpectralNeighborAnalysis2015,shapeevMomentTensorPotentials2016,zhangDeepPotentialMolecular2018} These so-called ML-IAPs typically express the PES as a function of local environment descriptors such as the interatomic distances and angles, or atomic densities, and have been demonstrated to significantly outperform classical IAPs across a broad range of chemistries.\cite{zuoPerformanceCostAssessment2020} Message-passing and graph deep learning models\cite{schuttSchNetContinuousfilterConvolutional2017,klicperaDirectionalMessagePassing2020,haghighatlariNewtonNetNewtonianMessage2021} have also been shown to yield highly accurate predictions of energies and/or forces of molecules as well as a limited number of crystals, such as  \ce{Li7P3S11}\cite{parkAccurateScalableGraph2021} and \ce{Li_xSi_y}\cite{cheonCrystalStructureSearch2020} for lithium-ion batteries. Nevertheless, no work has demonstrated a universally applicable IAP across the periodic table and for all crystal types.  

In the past decade, the advent of efficient and reliable electronic structure codes\cite{lejaeghereReproducibilityDensityFunctional2016} with high-throughput automation frameworks\cite{ongPythonMaterialsGenomics2013,jainFireWorksDynamicWorkflow2015, pizziAiiDAAutomatedInteractive2016,mathewAtomateHighlevelInterface2017} have led to the development of large federated databases of computed materials data, such as the Materials Project,\cite{jainCommentaryMaterialsProject2013}  AFLOW,\cite{curtaroloAFLOWLIBORGDistributed2012} Open Quantum Mechanical Database (OQMD),\cite{kirklinOpenQuantumMaterials2015a} NOMAD,\cite{draxlNOMADLaboratoryData2019} etc. Most of the focus has been on the utilization of the final outputs from the electronic structure computations carried out by these database efforts, namely, the equilibrium structures, energies, band structures and other derivative material properties, for the purposes of materials screening and design. Less attention has been paid to the huge quantities of PES data, i.e., intermediate structures and their corresponding energies, forces, stresses, that have been amassed in the process of performing structural relaxations.

In this work, we develop the formalism for a graph-based deep learning IAP by combining many-body features of traditional IAPs with those of flexible graph material representations. Utilizing the largely untapped dataset of more than 187,000 energies, 16,000,000 forces and 1,600,000 stresses from structural relaxations performed by the Materials Project since its inception in 2011, we trained a universal material graph with three-body interactions neural network (M3GNet) IAP for 89 elements of the periodic table with low energy, force, and stress errors. We demonstrate the applications of M3GNet in the calculations of phonon and elasticity, structural relaxations, etc. We further relaxed $\sim$30 millions of hypothetical structures for new materials discovery.

\section{Materials Graphs with Many-Body Interactions}

Mathematical graphs are a natural representation for crystals and molecules, with nodes and edges representing the atoms and the bonds between them, respectively. In traditional graph neural network (GNN) models for materials, information flows between the node, edge, and, optionally, state vector attributes via successive application of graph convolutional or update operations.\cite{xieCrystalGraphConvolutional2018, chenGraphNetworksUniversal2019, chenLearningPropertiesOrdered2021, decostAtomisticLineGraph2021} Typically, the input bond attribute is based on an interatomic pair distance measure, such as a Gaussian basis expansion. While such graph deep learning models have proven to be exceptionally effective for general materials property predictions,\cite{xieCrystalGraphConvolutional2018, chenGraphNetworksUniversal2019, chenLearningPropertiesOrdered2021,decostAtomisticLineGraph2021} they are not suitable as IAPs due to the lack of physical constraints such as continuity of energies and forces with changes with the length and number of bonds.

\begin{figure}
\centering
\includegraphics[width=0.99\textwidth]{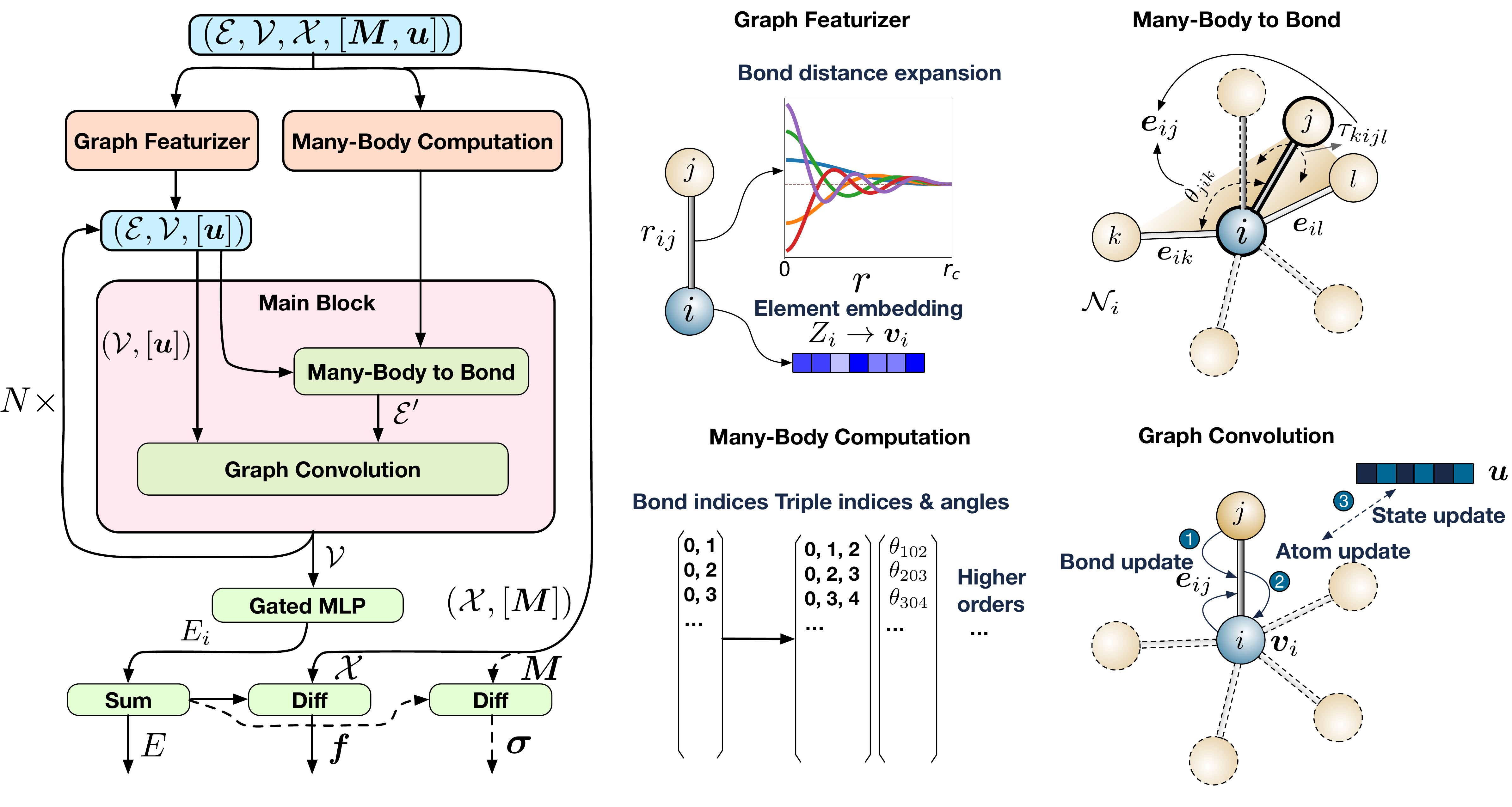}
\caption{\textbf{Schematic of the many-body graph potential and the major computational blocks.}  The model architecture starts from a position-included graph, and then goes through a featurization process, followed by main blocks, and the readout module with energy, force and stress outputs. The featurization process consists of the graph featurizer and the many-body computation module. In the graph featurizer, the atomic number of elements were embedded into a learnable continuous feature space, and the pair bond distances were expanded onto a basis set with values and derivatives up to second order going to zero at boundary. The many-body computation module calculates the three-body and many-body interaction atom indices and the associated angles. The main block consists of two main steps, namely the many-body to bond module and standard graph convolution.  The many-body to bond step calculates the new bond information $\bm{e}_{ij}$ by considering the full bonding environment $\mathcal{N}_i$ of atom $i$ via many-body angles such as $\theta_{jik}$, $\tau_{kijl}$, etc., and the bond length $r_{ik}, r_{ij}, r_{il}$, etc. The standard graph convolution updates bond, atom, and the optional state information iteratively. During the readout stage, atom information in the graph was passed to a gated MLP for obtaining atomic energy, which sums to the total energy. The derivatives of the total energy give force and stress outputs.}
\label{fig:schematic}
\end{figure}

Here, we develop a new materials graph architecture that explicitly incorporates many-body interactions (Figure \ref{fig:schematic}). The materials graph is represented as  $\mathcal{G} = (\mathcal{V}, \mathcal{E}, \mathcal{X}, [\bm{M}, \bm{u}]])$, where $\bm{v}_i \in \mathcal{V}$ is atom information for $i$, $\bm{e}_{ij} \in \mathcal{E}$ is the bond information for bond connected by atom $i$ and $j$, and $\bm{u}$ is the optional global state information, as temperature, pressure etc. A key difference with prior materials graph implementations is the addition of $\bm{x}_i \in \mathcal{X}$, the coordinates for atom $i$, and $\bm{M}$, the optional $3\times 3$ lattice matrix in crystals, which are necessary for obtaining tensorial quantities such as forces and stresses via auto-differentiation.

The neighborhood of atom $i$ is denoted as $\mathcal{N}_i$. Taking inspiration from traditional IAPs such as the Tersoff bond order potential,\cite{tersoffNewEmpiricalApproach1988} we consider all other bonds emanating from atom $i$ when calculating the bond interaction of $\bm{e}_{ij}$. To incorporate $n$-body interactions, each $\bm{e}_{ij}$ is updated using all distinct combinations of $n-2$ neighbors in the neighborhood of atom $i$ excluding atom $j$, i.e., $\mathcal{N}_i/j$, denoted generally as follows:
\begin{equation}
    \tilde{\bm{e}}_{ij} = \sum_{\substack{k_1, k_2, ..., k_{n-2} \in \mathcal{N}_i/j\\k_1 != k_2 != ... k_{n-2}}}\phi_n(\bm{e}_{ij}, \bm{r}_{ij}, \bm{v}_{j}, \bm{r}_{ik_1}, \bm{r}_{ik_2}, ..., \bm{r}_{ik_{n-2}}, \bm{v}_{k_1},  \bm{v}_{k_2}, ...,  \bm{v}_{k_{n-2}})
\end{equation}
where $\phi_n$ is the update function and $\bm{r}_{ik}$ is the vector pointing from atom $i$ to atom $k$. In practice, this $n$-body information exchange involves the calculation of distances, angles, dihedral angles, improper angles, etc., which escalates combinatorially with the order $n$ as $(M_i-1)!/(M_i-n+1)!$ where $M_i$ is the number of neighbors in $\mathcal{N}_i$. For brevity, we will denote this materials graph with $n$-body interactions neural network as M$n$GNet. In this work, we will focus on the incorporation of three-body interactions only, i.e., M3GNet.

Let $\theta_{jik}$ denote the angle between bonds $\bm{e}_{ij}$ and $\bm{e}_{ik}$. Here, we expand the three-body angular interactions using an efficient complete and orthogonal spherical Bessel function and spherical harmonics basis set, as proposed by \citet{klicperaDirectionalMessagePassing2020} The bond update equation can then be rewritten as:

\begin{eqnarray}
    \tilde{\bm{e}}_{ij} & = &  \sum_k j_l(z_{ln}\frac{r_{ik}}{r_c})Y_l^0(\theta_{jik}) \odot\sigma(\bm{W}_v\bm{v}_k + \bm{b}_v)f_c(r_{ij})f_c(r_{ik}) \label{eq:bonding_env} \\
    \bm{e}_{ij}^\prime & = & \bm{e}_{ij} + g(\tilde{\bm{W}_2} \tilde{\bm{e}}_{ij} + \tilde{\bm{b}}_2)\odot \sigma(\tilde{\bm{W}_1} \tilde{\bm{e}}_{ij} + \tilde{\bm{b}}_1) \label{eq:bond_update}
\end{eqnarray}
where $\bm{W}$ and $\bm{b}$ are learnable weights from the network, $j_l$ is the spherical Bessel function with the roots at $z_{ln}$, $r_c$ is the cutoff radius, $Y_l^0$ is the spherical harmonics function with $m=0$, $\odot$ is the element-wise product, $\sigma$ is the sigmoid activation function, $f_c(r) = 1-6(r/r_c)^5 + 15 (r/r_c)^4 - 10 (r/r_c)^3$ is the cutoff function ensuring the functions vanishes smoothly at the neighbor boundary,\cite{singraberLibraryBasedLAMMPSImplementation2019} and $g(x) = x\sigma(x)$ is the nonlinear activation function.\cite{ramachandranSearchingActivationFunctions2017}  $\tilde{\bm{e}}_{ij}$ is a vector of length $n_{max}l_{max}$ expanded by indices $l = {0, 1, ..., l_{max}-1}$ and $n = {0, 1, ..., n_{max} - 1}$. 

Following the $n$-body interaction update, several graph convolution steps are carried out sequentially to update the bond, atom and, optionally, state information, as follows:

\begin{eqnarray}
\bm{e}_{ij}^\prime &=& \bm{e}_{ij} +  \phi_e(\bm{v}_i\oplus\bm{v}_j\oplus\bm{e}_{ij}\oplus\bm{u}) \bm{W}_e^0\bm{e}_{ij}^0 \label{eq:gc_bond}\\
\bm{v}_i^\prime &=& \bm{v}_i + \sum_j\phi_e^\prime(\bm{v}_i \oplus \bm{v}_j \oplus \bm{e}_{ij}^\prime \oplus\bm{u}) \bm{W}_e^{0\prime}\bm{e}_{ij}^0 \label{eq:gc_atom}\\\
\bm{u}^\prime &=& g(\bm{W}^u_2g(\bm{W}^u_1(\frac{1}{N_v} \sum^{N_v}_{i} \bm{v}_i \oplus \bm{u}) + \bm{b}^u_1) + \bm{b}^u_2) \label{eq:gc_state}\
\end{eqnarray}
where $\phi_e(x)$ and $\phi_e^\prime(x)$ are gated multi-layer perceptrons as in Equation \ref{eq:gated}, $\oplus$ is the concatenation operator, $N_v$ is the number of atoms, and $\bm{e}_{ij}^0$ are the distance-expanded basis functions with values, first and second derivatives  smoothly go to zero at the cutoff boundary (see Methods). Such a design ensures that the target values and their derivatives up to second order change smoothly with changes in the number of bonds. $\bm{u}$ inputs and updates are optional to the models since not all structures or models have state attributes.

Each block of multi-step updates ($n$-body, bond, atom, state) can be repeated to construct models of arbitrary complexity, similar to previous materials graph network architectures.\cite{chenGraphNetworksUniversal2019}

\section{M3GNet Interatomic Potential}

To develop an IAP using the M3GNet architecture, crystal structures with corresponding energies ($E$), forces ($\bm{f}$) and stresses ($\bm{\sigma}$) as targets were used as training data. The model generate trainable targets via auto-differentiation with $\bm{f} = -\partial E/\partial \bm{x}$ and $\bm{\sigma} = V^{-1}\partial E/\partial \bm{\epsilon}$, where  $\bm{x}$ are the atomic coordinates, $V$ is the volume, and $\epsilon$ is the strain.

\subsection{Benchmark on IAP datasets}
As an initial benchmark, we selected a diverse DFT dataset of elemental energies and forces previously generated by \citet{zuoPerformanceCostAssessment2020} for fcc Ni, fcc Cu, bcc Li, bcc Mo, diamond Si and diamond Ge. From Table \ref{tab:main_benchmark_elements}, the M3GNet IAPs significantly outperform classical many-body potentials such as the embedded atom method (EAM) and modified EAM (MEAM) and performs comparably to local environment-based ML-IAPs such as the Behler-Parinello neural network potential (NNP)\cite{behlerGeneralizedNeuralNetworkRepresentation2007} and moment tensor potential (MTP)\cite{shapeevMomentTensorPotentials2016}. It should be noted that while ML-IAPs can achieve slightly lower energy and force errors than M3GNet IAPs, it comes at a substantial loss in flexibility in handling multi-element chemistries. Incorporating multiple elements in ML-IAPs results in a combinatorial explosion in number of regression coefficients and corresponding data requirements. For instance, the MTP requires $\mathcal{O}(n_{ele}^2)$ regression coefficients alone to describe the element interactions, where $n_{ele}$ is the number of elements. In contrast, the M3GNet architecture represents the elemental information for each atom (node) as a learnable embedding vector. Such a framework is readily extendable to multi-component chemistries. For instance, the M3GNet-all IAP trained on all six elements perform similarly to the M3GNet IAPs trained on individual elements. The M3GNet framework, like other GNNs, is able to capture long-range interactions without the need to increase the cutoff radius for bond construction (Figure S1). At the same time, unlike the previous GNN models, the M3GNet architecture still retains a continuous variation of energy, force and stress with changes of the number of bonds (Figure S2), a crucial requirement for IAPs.


\begin{table}[h!]
\begin{center}
\begin{adjustbox}{max width=\textwidth}
\caption{M3GNet models errors compared to the existing models EAM, MEAM, NNP, and MTP on the single-element dataset from Zuo et al.\cite{zuoPerformanceCostAssessment2020}. In each cell, the errors are reported in root mean squared error (RMSE) by averaging results from three independent model training. The M3GNet-all model trains all six elements in one model.} 
\label{tab:main_benchmark_elements}
\begin{tabular}{  L{2cm} | C{2cm}|C{2.5cm} | C{1.6cm} | C{1.6cm} | C{1.6cm} | C{1.6cm} } 
\hline
\hline
\textbf{Element} & \textbf{M3GNet} & \textbf{M3GNet-all} & \textbf{EAM} & \textbf{MEAM} & \textbf{NNP} & \textbf{MTP} \\
\hline
\multicolumn{7}{c}{{Energy ($10^{-3}$ eV atom$^{-1}$)}}\\
\hline
Ni & 0.9 & 1.9 & 8.5 & 23.0 & 2.3 & 0.8 \\
\hline
Cu & 1.8 & 2.3 & 7.5 & 10.5 &1.7 & 0.5  \\
\hline
Li & 2.5 & 4.7 & 368.6 & - &1.0 & 0.7 \\
\hline
Mo & 6.3 & 6.8 & 68.0 & 36.4 & 5.7 & 3.9 \\
\hline 
Si & 9.6 & 6.8 & - & 111.7 &9.9 & 3.0 \\
\hline
Ge & 9.4 & 5.9 &  - & - & 11.0 & 3.7 \\
\hline
\multicolumn{7}{c}{{Force ($10^{-3}$ eV Å$^{-1}$)}}\\
\hline
Ni & 37.4 & 37.0 & 110 &  330 &  67.3 &  26.9 \\
\hline
Cu & 17.0 & 16.9 & 120 & 240 & 63.0 &  13.5  \\
\hline
Li & 22.1 &  24.5 &  140 & - & 63.4 & 13.2 \\
\hline
Mo &  193.7 &  271.4 &  520 &  220 & 198.7 &  148.1 \\
\hline 
Si &  102.8 &  126.2 & - &  400 &174.2 &  88.1 \\
\hline
Ge &  76.4 & 78.4 &  - & - &  124.3 &  70.3 \\
\hline
\hline
\end{tabular}
\end{adjustbox}
\end{center}
\end{table}

\subsection{Universal Interatomic Potential for the Periodic Table} 

\begin{figure}
\centering
\includegraphics[width=0.9\textwidth]{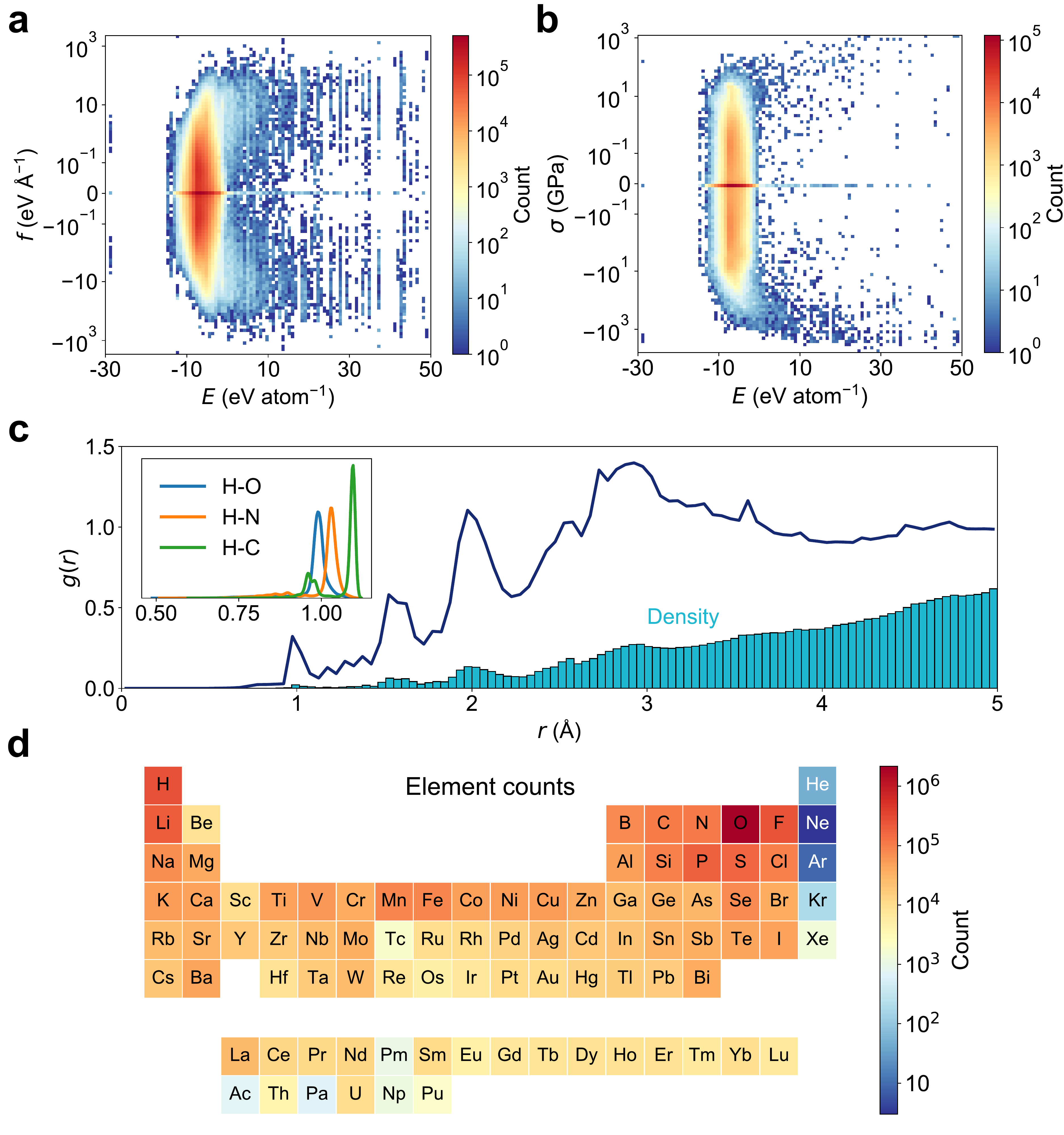}
\caption{\textbf{The distribution of the MPF.2021.2.8 dataset.} Structural energy per atom versus force components (\textbf{a}) and stress components (\textbf{b}) distributions. \textbf{c,} The radial distribution function $g(r)$ and pair atom distance distribution density. The short distance ($<$1.1 Å) density is made of mostly hydrogen bonded with O, C and N, illustrated in the inset. \textbf{d,} Element counts for all atoms in the dataset, covering 89 elements across the periodic table.}
\label{fig:mpff2021}
\end{figure}

To develop an IAP for the entire periodic table, we leveraged on one of the largest open databases of DFT crystal structure relaxations in the world - the Materials Project.\cite{jainCommentaryMaterialsProject2013} The Materials Project performs a sequence of two relaxation calculations\cite{mathewAtomateHighlevelInterface2017} with the Perdew-Burke Ernzerhof (PBE)\cite{perdewGeneralizedGradientApproximation1996} generalized gradient approximation (GGA) functional or the GGA+U method\cite{anisimovBandTheoryMott1991} for every unique input crystal, typically obtained from an experimental database such as the Inorganic Crystal Structure Database (ICSD).\cite{hellenbrandtInorganicCrystalStructure2004} Our initial dataset comprises a sampling of the energies, forces and stresses from the first and middle ionic steps of the first relaxation and the last step of the second relaxation for calculations in the Materials Project database that contains ``GGA Structure Optimization'' or ``GGA+U Structure Optimization'' task types as of Feb 8, 2021. The snapshots that have a final energy per atom greater than 50 eV atom$^{-1}$ or atom distance less than 0.5 Å were excluded, since those tend to be the result of errors in the initial input structure. In total, this ``MPF.2021.2.8'' dataset contains 187,687 ionic steps of 62,783 compounds, with 187,687 energies, 16,875,138 force components, and 1,689,183 stress components. The dataset covers an energy, force and stress range of [-28.731, 49.575] eV atom$^{-1}$, [-2570.567, 2552.991] eV Å$^{-1}$ and [-5474.488, 1397.567] GPa, respectively (Figure \ref{fig:mpff2021}a,b). The majority of structures have formation energies between -5 and 3 eV atom$^{-1}$, as shown in Figure S3. While the distribution of forces is relatively symmetric, the stress data contains a slightly higher proportion of negative (compressive) stresses than positive stresses due to the well-known tendency of the PBE functional to underbind. The radial distribution function $g(r)$ (Figure \ref{fig:mpff2021}c) shows that the dataset also spans a broad range of interatomic distances, including small distances of less than 0.6 Å that are essential for the M3GNet model to learn the repulsive forces at close distances. The dataset encompasses 89 elements of the periodic table. More information about the MPF.2021.2.8 data distribution is provided in Table S1. This dataset is then split into the training, validation and test data in the ratio of 90\%, 5\% and 5\%, respectively, according to materials not data points. Three independent data splits were performed.

In principle, an IAP can be trained on only energies, or a combination of energies and forces. In practice, the M3GNet IAP trained only on energies (M3GNet-$E$) was unable to achieve reasonable accuracies for predicting either forces or stresses, with mean absolute errors (MAEs) great than even the mean absolute deviation of the data (Table S2). This is the result of the amplification of errors when calculating the derivatives when only energy data is used. The M3GNet models trained with energies + forces (M3GNet-$EF$) and energies + forces + stresses (M3GNet-$EFS$) achieved relatively similar energy and force MAEs, but the MAE in stresses of the M3GNet-$EFS$ was about half that of the M3GNet-$EF$ model. Accurate stress predictions are necessary for applications that involve lattice changes, such as structural relaxations or $NpT$ molecular dynamics (MD) simulations. Our results suggest that it is critical to include all three properties (energy, force, and stress) in the model training to obtain a practical IAP. The final M3GNet-$EFS$ IAP (henceforth, referred to simple as the M3GNet model for brevity) achieved an average of 0.035 eV atom$^{-1}$, 0.072 eV Å$^{-1}$, and 0.41 GPa for energy, force, and stress test MAE, respectively.

\begin{figure}
\centering
\includegraphics[width=0.97\textwidth]{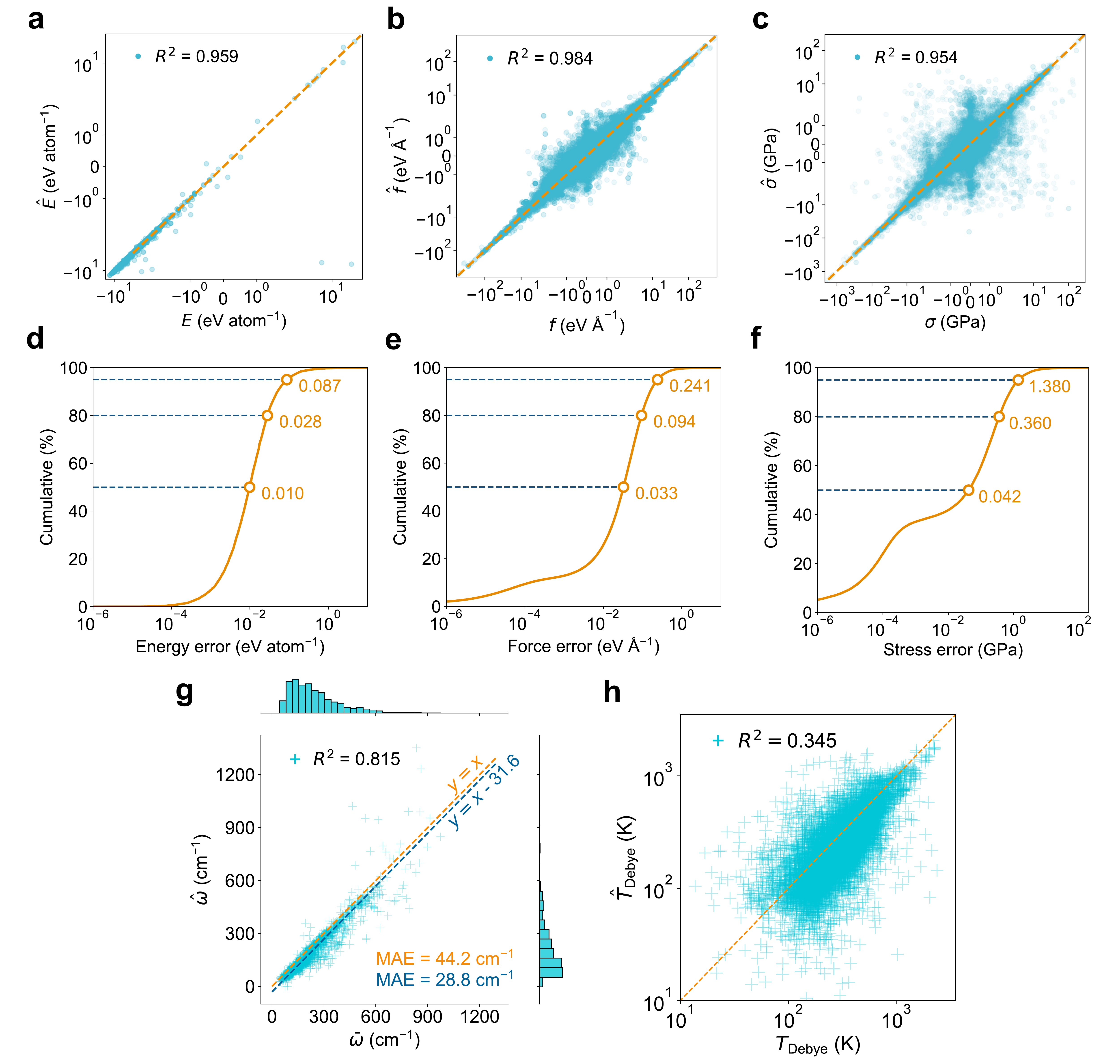}
\caption{\textbf{The model predictions on the test dataset compared to DFT calculations.} \textbf{a-c}, The parity plots for energy, force and stress, respectively. The model predicted results are $\hat{E}, \hat{f}$, and $\hat{\sigma}$. The dashed lines are $y=x$ as guides for the eye. \textbf{d-e}, The cumulative distribution of errors for energy, force and stress, respectively. The horizontal dashed lines indicate the model errors, from bottom to top, at 50\%, 80\% and 95\%. \textbf{g,} The comparison between model calculated 1,521 phonon density of state (DOS) center data ($\hat{\omega}$) and the PBEsol DFT calculations ($\bar{\omega}$) from Petretto et al.\cite{petrettoHighthroughputDensityfunctionalPerturbation2018a} and \textbf{h,} the 11,848 Debye temperatures (excluding negative moduli) calculated from M3GNet model ($\hat{T}_{\rm{Debye}}$) and PBE DFT elastic tensors from de Jong et al.\cite{dejongChartingCompleteElastic2015a}}
\label{fig:mpff2021_prediction}
\end{figure}

We further investigated the test error distributions of one final M3GNet model. Generally, the model predictions and the DFT ground truth match well as revealed by the high linearity and the $R^2$ values for the linear fitting between DFT and model predictions (Figure \ref{fig:mpff2021_prediction}a-c) The cumulative distribution of the model errors indicate that 50\% of the data has energy, force, and stress errors lower than 0.01 eV atom$^{-1}$, 0.033 eV Å$^{-1}$ and 0.042 GPa, respectively (Figure \ref{fig:mpff2021_prediction}d-f). Even more stringent tests were carried out using phonon and elasticity calculations, which were not part of the original training data. The M3GNet model can reproduce accurate phonon dispersion curves and density of states (DOS) of $\beta$-cristobalite, stishovite, and $\alpha$-quartz \ce{SiO2} (Figure S4) to quantitative agreements with expensive DFT computations.\cite{petrettoHighthroughputDensityfunctionalPerturbation2018a}. The M3GNet phonon DOS centers $\bar\omega$ from phonon calculations using predicted forces and the frozen phonon approach are also in good agreement with DFT computed values with a MAE of 44.2 cm$^{-1}$ (Figure \ref{fig:mpff2021_prediction}g).\cite{petrettoHighthroughputDensityfunctionalPerturbation2018a} The systematic underestimation by the M3GNet model relative to DFT is likely due to the different choices of pseudopotentials; the DFT phonon calculations were performed using the PBEsol\cite{perdewRestoringDensityGradientExpansion2008} functional while the M3GNet training data comprised of PBE/PBE+U calculations.\cite{kresseInitioMolecularDynamics1993,kresseEfficiencyAbinitioTotal1996} This systematic underestimation can be corrected with a constant shift of 31.6 cm$^{-1}$ and the MAE reduces to 28.8 cm$^{-1}$. Such errors are even lower than a state-of-the-art phonon DOS peak position prediction model which reported MAE of 36.9 cm$^{-1}$.\cite{dunnBenchmarkingMaterialsProperty2020a} We note that the DOS peak prediction model does not exhibit a systematic shift as it was directly fitted on the data by minimizing a mean-squared error. Similar to DFT, the relationship $\bar\omega \propto 1/(\overline m)^2$, where $\overline{m}$ is the average atomic mass, is obtained (Figure S5). The M3GNet-calculated Debye temperatures are less accurate (Figure \ref{fig:mpff2021_prediction}h), which can be attributed to relative poor M3GNet predictions of the shear moduli ($R^2$ = 0.134) (Figure S6), though the the bulk moduli predictions ($R^2$ = 0.757) are reasonable.

\begin{figure}
\centering
\includegraphics[width=0.97\textwidth]{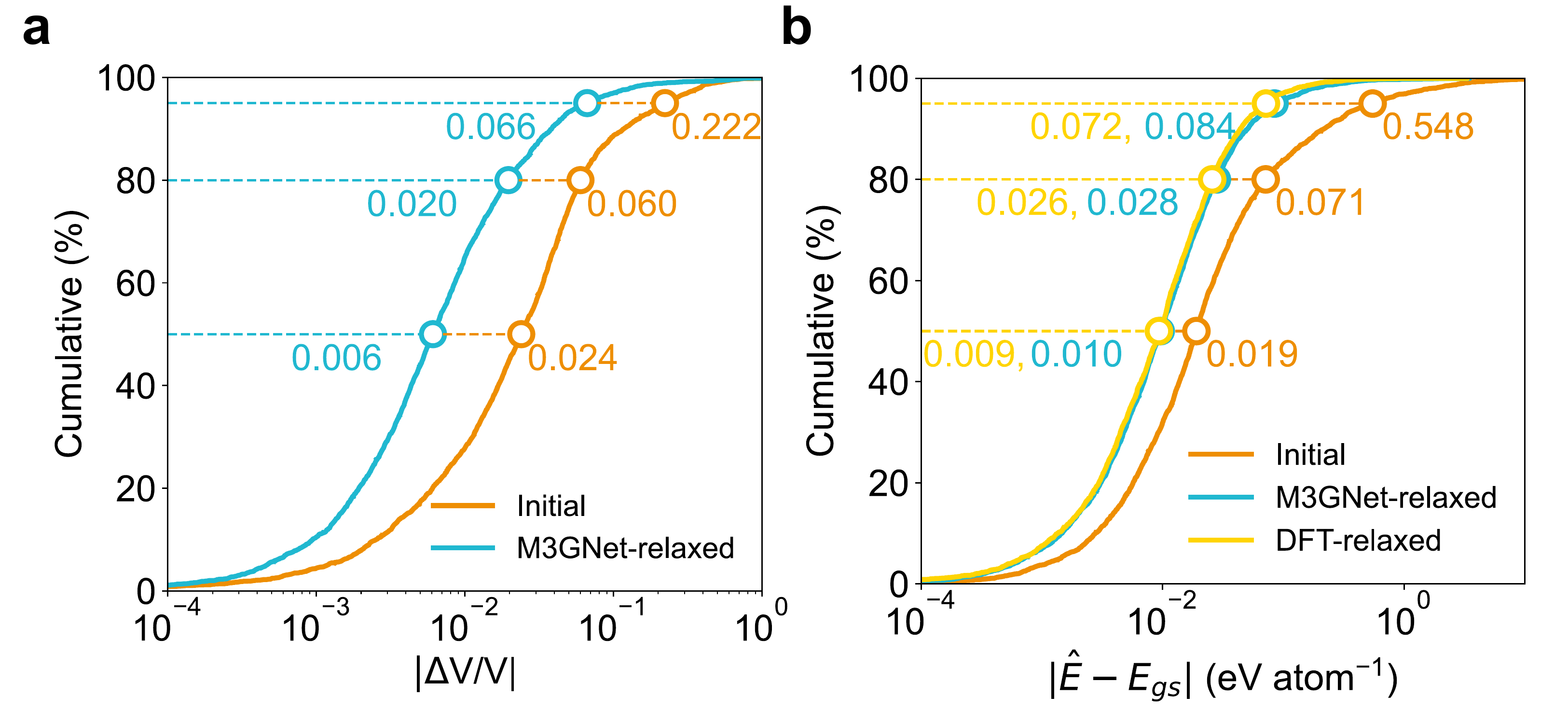}
\caption{\textbf{Relaxation of crystal structures with M3GNet}. \textbf{a}, Distribution of the absolute percentage error in volumes of M3GNet-relaxed structures relative to DFT-relaxed structures. \textbf{b}, The differences between M3GNet-predicted energies $\hat{E}$ and ground state (gs) energies $E_{gs}$ using the initial, M3GNet-relaxed and DFT-relaxed structures. $E_{gs}$ is defined as the DFT energy of the DFT-relaxed crystal. The horizontal lines mark the 50th, 80th, and 95th percentiles of the distributions and the corresponding $x$ axis values are annotated. }
\label{fig:relax_overall}
\end{figure}

The M3GNet model was then applied in a simulated materials discovery workflow where the final DFT structures are not known \textit{a priori}. M3GNet relaxations were carried out on the initial structures from the test dataset of 3,140 materials. M3GNet relaxation yields crystals that have volumes much closer to the DFT reference volumes (Figure \ref{fig:relax_overall}a). While 50\% and 5\% of the initial input structures have volumes that differ from the final DFT relaxed crystals by more than 2.4\% and 22.2\%, respectively, these errors are reduced to 0.6\% and 6.6\% via M3GNet relaxation. Correspondingly, the errors in the predicted energies $\hat{E}$ are also much smaller (Figure \ref{fig:relax_overall}b). Using the initial structures for direct model predictions, the energy differences distribute broadly, with considerable amount of structures having errors larger than 0.1 eV atom$^{-1}$.  All errors here were calculated relative to the DFT energies of the final DFT-relaxed structures for each material. The overall MAE is 0.169 eV atom$^{-1}$ with $\sim 20\%$ of the structures having errors greater than 0.071 eV atom$^{-1}$ (Figure \ref{fig:relax_overall}b). These errors are far too large for reliable estimations of materials stability, given that 90\% of all inorganic crystals in the ICSD has an energy above the convex hull of less than 0.067 eV atom$^{-1}$.\cite{sunThermodynamicScaleInorganic2016} In contrast, energy calculations on the M3GNet-relaxed structures yield a MAE of 0.035 eV atom$^{-1}$ and 80\% of the materials have errors less than 0.028 eV atom$^{-1}$. The error distributions using M3GNet relaxed structures are close to the case where we know the DFT final structures, as shown in Figure \ref{fig:relax_overall}b, suggesting that M3GNet potential can be accurate in helping getting the correct structures. In general, relaxations with M3GNet converges rapidly, as shown in Figure S7. An example of M3GNet relaxation is shown in Figure S8 for \ce{K57Se34} (mp-685089), a material with one of the largest energy change during relaxation. Convergence is achieved after about 100 steps when the forces falls under 0.1 eV Å$^{-1}$. The X-ray diffraction (XRD) pattern of the M3GNet-relaxed structure also resembles the counterpart from DFT relaxation (Figure  S8g). This relaxation can be performed on a laptop in about 22 seconds on a single CPU core of Intel(R) Xeon(R) CPU E5-2620 v4 @ 2.10GHz, while the corresponding DFT relaxation took 15 hours on 32 cores in the original Materials Project calculations.

\section{New Materials Discovery}

The ability of M3GNet to accurately and rapidly relax arbitrary crystal structures and predict their energies makes it ideal for large-scale  materials discovery. To generate hypothetical materials, combinatorial isovalent ionic substitutions based on the common oxidation states of non-noble-gas element were performed on 5,283 binary, ternary and quaternary structural prototypes in the 2019 version of the ICSD\cite{hellenbrandtInorganicCrystalStructure2004} database. Only prototypes with less than 51 atoms were selected for computational speed considerations. Further filtering was performed to exclude structures with non-integer or zero-charged atoms. A total of 31,664,858 hypothetical materials candidates were generated, more than 200 times the total number of unique crystals in the Materials Project today. All structures were relaxed using the M3GNet model and their signed energy distance to the Materials Project convex hull were calculated using the M3GNet IAP-predicted energy ($E_{\rm{hull-m}}$). We acknowledge that some of the generated structures may compete with each other for stability. However, to avoid introducing additional uncertainties into the $E_{\rm{hull-m}}$ predictions, we have elected to compute $E_{\rm{hull-m}}$ relative to ground-truth DFT energies in the Materials Project as opposed to the higher uncertainty M3GNet-computed energies. A zero or negative $E_{\rm{hull}}$ means that the material is predicted to be potentially stable compared to known materials in MP. The more negative the $E_{\rm{hull}}$, the greater the probability that a material is likely to be stable after accounting for uncertainties in the M3GNet-predicted energies. In total, 1,849,096 materials have $E_{\rm{hull-m}}$ less than 0.001 eV atom$^{\rm{-1}}$. We then excluded materials that have non-metal ions in multiple valence states, e.g., materials containing \ce{Br+} and \ce{Br-} at the same time, etc. It is well-known that PBE overbinds single-element molecules such as \ce{O2}, \ce{S8}, \ce{Cl2}, etc. and negative anion energy corrections are applied to ionic compounds in Materials Project to offset such errors.\cite{wangOxidationEnergiesTransition2006} However the corrections are based mostly on composition, which may artificially over-stabilize materials with multi-valence non-metal ions. We have developed a searchable database for the generated hypothetical structures and their corresponding M3GNet-predicted properties at http://matterverse.ai.

\begin{figure}
\centering
\includegraphics[width=0.99\textwidth]{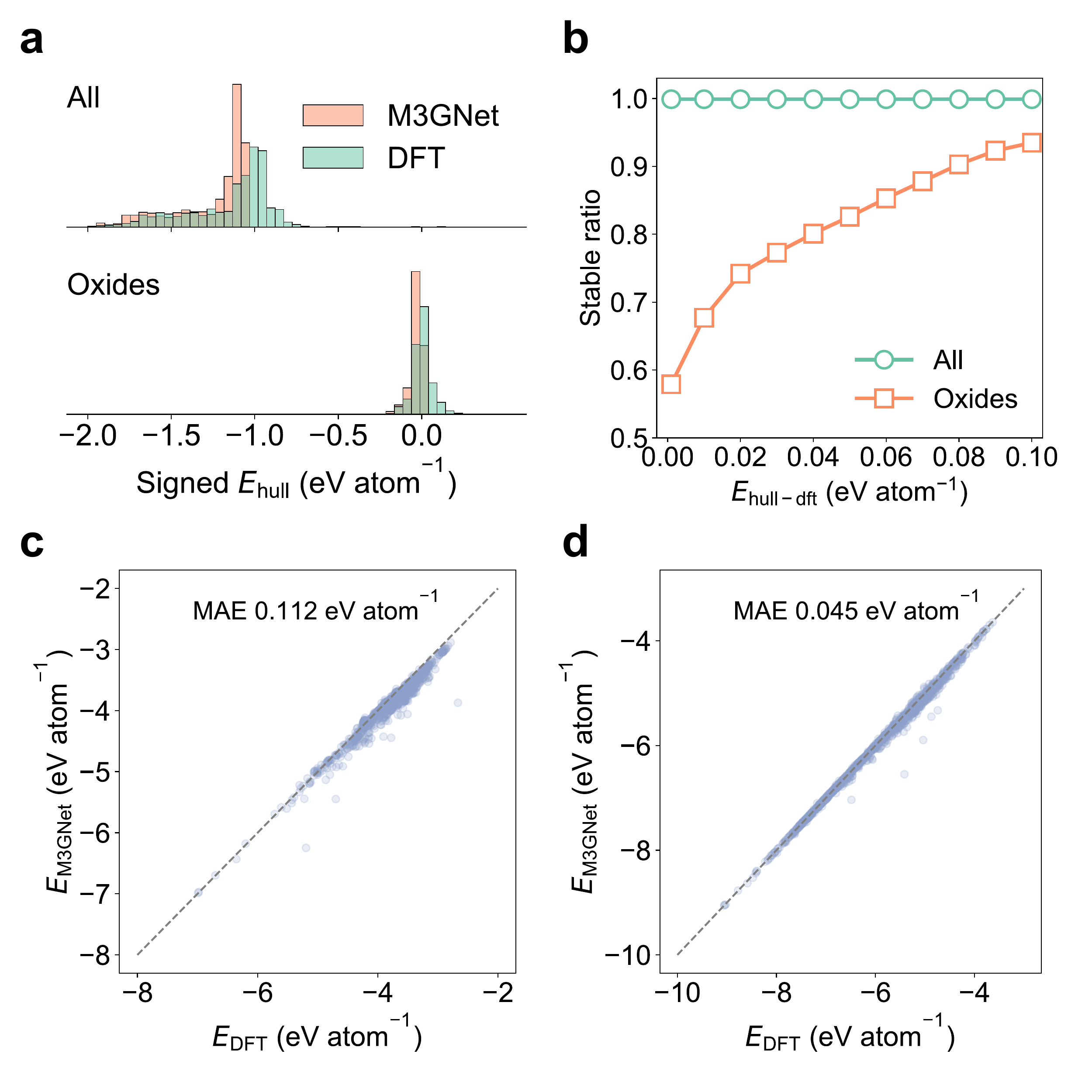}
\caption{\textbf{Discovery of stable materials using M3GNet.} \textbf{a,} The signed $E_{\rm{hull}}$ distribution for the top-1000 lowest $E_{\rm{hull-m}}$ materials from any chemistry (All) and oxides only (Oxides). \textbf{b,} Fraction of materials below $E_{\rm{hull}-dft}$ among top-1000 materials in the All and Oxides categories. \textbf{c, d}, Plot of the final M3GNet predicted energy against final DFT energy for the (c) All and (d) Oxides categories.}
\label{fig:stable_materials}
\end{figure}

\begin{figure}
\centering
\includegraphics[width=0.99\textwidth]{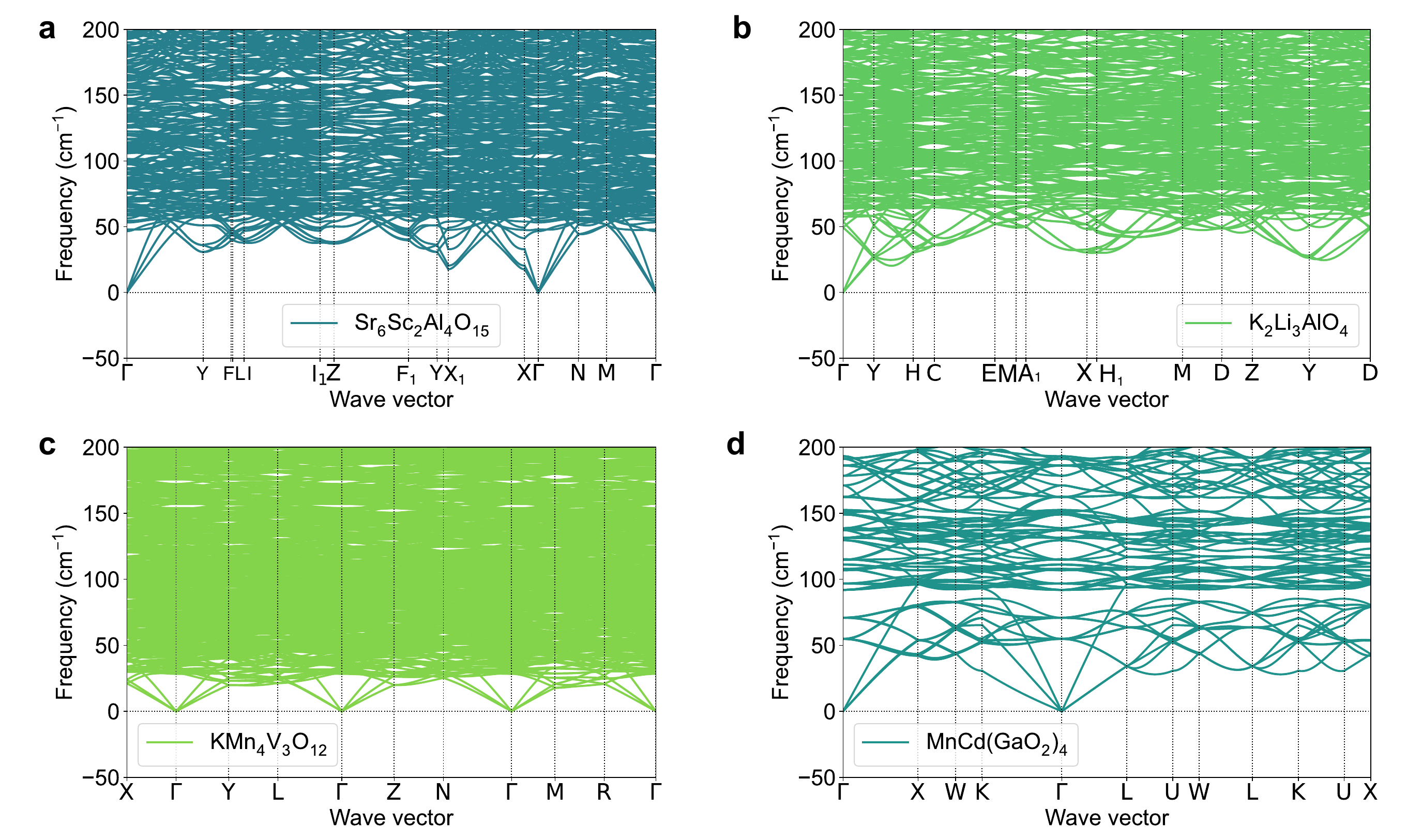}
\caption{\textbf{M3GNet-calculated phonon dispersion curves of four new materials predicted to be thermodynamically and dynamically stable.} \textbf{a,} \ce{Sr6Sc2Al4O15}; \textbf{b,} \ce{K2Li3AlO4}; \textbf{c,} \ce{KMN4V2O12}; \textbf{d,} \ce{MnCd(GAO2)4}}
\label{fig:phonon_all}
\end{figure}

A formation energy model based on the Matbench\cite{dunnBenchmarkingMaterialsProperty2020a} Materials Project data was developed \rev{using the same architecture as the M3GNet IAP model (see Table S3)}. Materials with a difference in the signed energy distance to the Materials Project convex hull from this model ($E_{\rm{hull-f}}$) and $E_{\rm{hull-m}}$ greater than 0.2 eV atom$^{\rm{-1}}$ were then discarded in the subsequent DFT analysis. \rev{This additional step removes materials with higher energy prediction uncertainties, which account for 13.1\% (243,820) of the predicted materials. It should be noted that this step can also be omitted to simplify the discovery workflow, though potentially with impact on the hit rate of stable materials discovery.} The top-1000 lowest $E_{\rm{hull-m}}$ materials from any chemistry as well as the top-1000 metal oxides with elements from the first five rows (excluding Tc due to radioactivity and Rb due to high dominance) were then selected for validation via DFT relaxation and energy calculations. Only the most stable polymorphs were selected for each composition. It was found that the distribution in the DFT calculated $E_{\rm{hull}-dft}$ matches well with the distributions of $E_{\rm{hull-m}}$ (Figure \ref{fig:stable_materials}a). For most computational materials discovery efforts, a positive threshold, typically around 0.05-0.1 eV atom$^{-1}$, is applied to identify synthesizable materials. This positive threshold accounts for both errors in DFT calculated energies as well as the fact that some thermodynamically meta-stable materials can be realized experimentally. Of the top-1000 materials from any chemistry, 999 were found to have a $E_{\rm{hull}-dft} < 0.001$ eV atom$^{-1}$ (Figure \ref{fig:stable_materials}b) and none of them were in the Materials Project database. For the top-1000 oxides, 579, 826, and 935 were found to be synthesizable based on $E_{\rm{hull}-dft}$ thresholds of 0.001, 0.05 and 0.1 eV atom$^{-1}$, respectively (Figure \ref{fig:stable_materials}b). Out of the 579 DFT-stable oxides, only five, namely \ce{Mg4Nb2O9}, \ce{Sr3V2O8}, \ce{K2SnO2}, \ce{Cd(RhO2)2}, \ce{CoMnO4}, were previously known and matched with the Materials Project structures. The effectiveness of the M3GNet IAP relaxations can be seen in Figure S9, which show that the energy changes during subsequent DFT relaxations (of the MEG3Net-relaxed structures) are at least one order of magnitude smaller than the energy changes during M3GNet relaxation. The final M3GNet-relaxed energies are in excellent agreement with the final DFT-relaxed energies, with MAEs of 0.112 and 0.045 eV atom$^{-1}$ for the top 1000 materials in any chemistry and the oxide chemistry, respectively (Figures \ref{fig:stable_materials}c-d). Using the M3GNet IAP, we have also assessed the dynamic stability of the 1578 materials with $E_{\rm{hull}-dft} < 0.001$ eV atom$^{-1}$ using phonon calculations. A total of 328 materials do not exhibit imaginary frequencies in their M3GNet phonon dispersion curves. Four phonon dispersion curves are shown in Figure \ref{fig:phonon_all}. The others are provided in Data Availability.

As an additional evaluation of the performance of M3GNet for materials discovery, we computed the discovery rate, i.e., the fraction of DFT-stable materials ($E_{\rm{hull}-dft} \leq 0$) for 1000 structures uniformly sampled from the $\sim$ 1.8 million materials with $E_{\rm{hull-m}} < 0.001$ eV atom$^{\rm{-1}}$. The discovery rate remains close to 1.0 up to a $E_{\rm{hull-m}}$ threshold of around 0.5 eV atom$^{-1}$ and remains at a reasonably high value of 0.31 at the strictest threshold of 0.001 eV atom$^{-1}$, as shown in Figure S10. For this material set, we also compared the DFT relaxation time cost with and without M3GNet pre-relaxation. The results show that without M3GNet pre-relaxation, the DFT relaxation time cost is about three times of that with the M3GNet relaxation, as shown in Figure S11.

\section{Discussion}

A universal IAP such as M3GNet has applications beyond crystal structure relaxation and stability predictions. For instance, a common application of IAPs is in molecular dynamics (MD) simulations to obtain transport properties such as diffusivity and ionic conductivity. The Arrhenius plot from MD simulations at multiple temperatures of the recently discovered Li superionic conductor \ce{Li3YCl6} is shown in Figure S12. The results agree well with the ionic conductivity and activation barriers from previous \textit{ab initio} MD simulations.\cite{wangLithiumChloridesBromides2019} Training an IAP for a complex multi-component systems such as \ce{Li3YCl6} is typically a highly-involved process,\cite{qiBridgingGapSimulated2021} while the M3GNet IAP can be universally applied to any material without further retraining. As shown in Figure S13, M3GNet MD calculations can potentially be applied to a wide range of Li-containing compounds to identify potential lithium superionic conductors. Furthermore, the M3GNet IAP can also serve as a surrogate model in lieu of DFT with other structural exploration techniques such as evolutionary algorithms like USPEX\cite{glassUSPEXEvolutionaryCrystal2006} and CALYPSO\cite{wangCALYPSOMethodCrystal2012} or generative models such as CDVAE\cite{xieCrystalDiffusionVariational2021} to generate more diverse and unconstrained candidates.

It should be noted that the current M3GNet IAP reported in this work is merely the best that can be done at present with available data. Further improvements in accuracy can be achieved through several efforts. First, the training data for the M3GNet IAP comes from DFT relaxation calculations in the Materials Project, which were performed with less stringent convergence criteria such as a lower energy cutoff and sparser $k$-point grids. For IAP development, a best practice is to obtain accurate energies, forces and stresses via single-point, well-converged DFT calculations for training data. Building such a database is an extensive effort that is planned for future developments in the Materials Project. Second, active learning strategies, for instance, by using the DFT relaxation data from the M3GNet-predicted stable crystals in a feedback loop, can be used to systematically improve the M3GNet IAP, especially in under-explored chemical spaces with the greatest potential for novel materials discoveries. Nevertheless, about 1.8 million of the 31 million candidates were predicted to be potentially stable or meta-stable by M3GNet against materials in the Materials Project, which already expands the potential exploration pool by an order of magnitude over the $\sim$140,000 crystals in the Materials Project database today. We shall note that the potentially stable materials will need to be further verified with DFT calculations and experimental synthesis.

The model uncertainty will also play a role in further decreasing the number of true discoveries. The candidate space contains 294,643 chemical systems, while the Materials Project has only about 47,000 chemical systems. Hence, we likely see many extrapolations in the predictions. For example, some of the most uncertain predictions are from chemical systems such as F-Fe-Se-W, F-Fe-S-W, F-Fe-Te-W, F-Ni-S-W, and Co-F-S-W, which are not represented in the Materials Project training data.

Finally, the M3GNet framework is not limited to crystalline IAPs or even IAPs in general. The M3GNet formalism without lattice inputs and stress outputs is naturally suited for molecular force fields. When benchmarked on MD17 and MD17-CCSD(T) molecular force field data (Table S4 and S5),\cite{chmielaMachineLearningAccurate2017,schuttQuantumchemicalInsightsDeep2017,chmielaExactMolecularDynamics2018} the M3GNet models were found to be more accurate than the embedded atom neural network (EANN) force field\cite{zhangEmbeddedAtomNeural2019} and perform similarly to the state-of-the-art message-passing networks and equivariant neural network models. Moreover, by changing the readout section from summed atomic energy as in Figure \ref{fig:schematic} to intensive property readout, the M3GNet framework can be used to develop surrogate models for property prediction. We trained M3GNet models on the Matbench materials data covering nine general crystal materials properties (Table S3)\cite{dunnBenchmarkingMaterialsProperty2020a}. In all cases, the M3GNet models achieved excellent accuracies.

\textbf{Correspondence} Correspondence and requests for materials should be addressed to C.C.~(email: chenc273@outlook.com) or S.P.O.~(email: ongsp@eng.ucsd.edu).

\begin{acknowledgement}
This work was primarily supported by the Materials Project, funded by the U.S. Department of Energy, Office of Science, Office of Basic Energy Sciences, Materials Sciences and Engineering Division under contract no. DE-AC02-05-CH11231: Materials Project program KC23MP. The lithium superionic conductor analysis portion of the work was funded by the LG Energy Solution through the Frontier Research Laboratory (FRL) Program. This work used the Extreme Science and Engineering Discovery Environment (XSEDE), which is supported by National Science Foundation grant number ACI-1548562.

\end{acknowledgement}

\section{Author contributions} C.C. and S.P.O. conceived the idea and designed the work. C.C. implemented the models and performed the analysis. C.C. and S.P.O. wrote the manuscript and contributed to the discussion and revision.

\section{Ethics Declaration} 
\subsection{Competing Interests }
The authors declare that they have no competing financial interests.

\section{Methods}

\subsection{Model construction}
\subsubsection{Neural network definition}

If we denote one layer of the perceptron model as 
\begin{equation}
    \mathcal{L}_g^k : x \mapsto g(\bm{W}_k x + \bm{b}_k)
\end{equation}
then the $K$-layer multi-layer perceptron (MLP) can be expressed as 
\begin{equation}
    \xi_K(x) = (\mathcal{L}_g^{K} \circ \mathcal{L}_g^{K-1} \circ ... \mathcal{L}_g^1) (x) \label{eq:mlp}
\end{equation}
The $K$-layer gated MLP becomes
 \begin{equation}
 \phi_K(x) = ((\mathcal{L}_g^{K} \circ \mathcal{L}_g^{K-1} \circ ... \mathcal{L}_g^1) (x)) \odot ((\mathcal{L}_\sigma^{K} \circ \mathcal{L}_g^{K-1} \circ ... \mathcal{L}_g^1) (x))  \label{eq:gated}
 \end{equation} 
 where $\mathcal{L}_\sigma^K(x)$ replaces the activation function $g(x)$ of  $\mathcal{L}_g^K(x)$  to sigmoid function $\sigma(x)$ and $\odot$ denotes element-wise product. The gated MLP consists of the normal MLP before $\odot$ and the gate network after $\odot$. 

\subsubsection{Model architecture}
Materials graphs were constructed using a radial cutoff of 5 Å. For computational efficiency considerations, the three-body interactions were limited to within a cutoff of 4 Å. The graph featurizer converts the atomic number into embeddings of dimension 64. The bond distances were expanded using the continuous and smooth basis function proposed by Kocer et al.\cite{kocerNovelApproachDescribe2019}, which ensures that the first and second derivatives vanish at the cutoff radius. 

\begin{equation}
    h_m(r) = \frac{1}{\sqrt{d_m}}\left[f_m(r) + \sqrt{\frac{e_m}{d_{m-1}}}h_{m-1}(r)\right] \label{eq:gn}
\end{equation}

where 
\begin{eqnarray}
    d_m & = & 1 - \frac{e_m}{d_{m-1}} \\
    e_m & = & \frac{m^2(m+2)^2}{4(m+1)^4 + 1}  \\
    f_m(r) & = & (-1)^m \frac{\sqrt{2}\pi}{r_c^{3/2}} \frac{(m+1)(m+2)}{\sqrt{(m+1)^2 + (m+2)^2}} \left(sinc\left(r\frac{(m+1)\pi}{r_c}\right) + sinc \left(r \frac{(m+2)\pi}{r_c}\right)\right)\\
    sinc(x) & = & \frac{\sin{x}}{x} 
\end{eqnarray}

$\bm{e}_{ij}^0$ is a vector formed by $m$ basis functions of $h(r)$. 

\begin{equation}
    \bm{e}_{ij}^0 (r_{ij}) = [h_1(r_{ij}), h_2(r_{ij}), ..., h_m(r_{ij})]
\end{equation}
In this work, we used three basis functions for the pair distance expansion.

The main blocks consist of three three-body information exchange and graph convolutions ($N=3$ in Figure \ref{fig:schematic}). By default, the $\bm{W}$'s and $\bm{b}$'s in the perceptron model gives output dimensions of 64. Each gated MLP ($\phi_e(x)$ and $\phi_e^\prime(x)$ in Equations \ref{eq:gc_bond} and \ref{eq:gc_atom}) have two layers with 64 neurons in each layer. 

For the prediction of extensive properties such as total energies, a three-layer gated MLP (Equation \ref{eq:gated}) was used on the atom attributes after the graph convolution and sum the outputs as the final prediction, i.e.,

\begin{equation}
    p_{\rm{ext}} = \sum_i \phi_3(\bm{v}_i) 
\end{equation}
The gated MLP $\phi_3(x)$ has a layer neuron configuration of [64, 64, 1] and no activation in the last layer of the normal MLP part. 

For the prediction of intensive properties, the readout step was performed as follows after the main blocks.

\begin{equation}
    p_{\rm{int}} = \xi_3(\sum_i w_i \xi_2(\bm{v}_i) \oplus \bm{u})
\end{equation}
with weights $w_i$ summing to 1 and defined as 
\begin{equation}
    w_i = \frac{\xi_3^\prime(\bm{v}_i)}{\sum_i \xi_3^\prime(\bm{v}_i)} 
\end{equation}
$\xi_3$ and $\xi_3^\prime$ have neuron configurations of [64, 64, 1] to ensure the output is scalar. For regression targets, there is no activation in the final layer of MLP, while for classification targets, the last layer activation is chosen as the sigmoid function. 

In the training of MPF.2021.2.8 data, the M3GNet model contains three main blocks with 227,549 learnable weights.

\subsection{Model training}

The Adam optimizer\cite{kingmaAdamMethodStochastic2017} was used with initial learning rate of 0.001, with a cosine decay to 1\% of the original value in 100 epochs. During the optimization, the validation metric values were used to monitor the model convergence, and training was stopped if the validation metric did not improve for 200 epochs. For the elemental IAP training, the loss function was the mean squared error (MSE). For other properties, the Huber loss function\cite{huberRobustEstimationLocation1964} with $\delta$ set to 0.01 was used. For the universal IAP training, the total loss function includes the loss for energy, forces, and, in inorganic compounds, also the stresses. Batch size of 32 was used in model training.

\begin{equation}
    L = \ell(e, e_D) + w_f \ell(\bm{f}, \bm{f_D}) + w_\sigma \ell(\bm{\sigma}, \bm{\sigma_D})
\end{equation}
where $\ell$ is the Huber loss function, $e$ is energy per atom, $\bm{f}$ is the force vector, $\bm{\sigma}$ is the stress, and $w$'s are the scalar weights. The subscript $D$ indicates data from DFT. 

Before M3GNet IAP fitting, we fit the elemental reference energies using linear regression of the total energies. We first featurize a composition into a vector $c = [c_1, c_2, c_3, ..., c_{89}]$ where $c_i$ is the number of atoms in the composition that has the atomic number $i$. The composition feature vector $c$ is mapped to the total energy of the material $E$ via $E = \sum_i c_i E_i$, where $E_i$ is the reference energy for element with atomic number $i$ that can be obtained by linear regression of the training data. Then, the elemental reference energies were subtracted from the total energies to improve M3GNet model training stability. We set $w_f = 1$ and $w_\sigma = 0.1$ during training the MPF.2021.2.8 data.

\subsection{Software implementation}

The M3GNet framework was implemented using the Tensorflow\cite{abadiTensorFlowSystemLargeScale2016} package. All crystal and molecular structure processing were performed using the Python Materials Genomics (pymatgen)\cite{ongPythonMaterialsGenomics2013} package. The structural optimization was performed using the FIRE\cite{bitzekStructuralRelaxationMade2006} algorithm implemented in the atomic simulation environment (ASE)\cite{larsenAtomicSimulationEnvironment2017}. The MD simulations were performed in the \textit{NVT} ensemble using ASE.\cite{larsenAtomicSimulationEnvironment2017} Phonon calculations were performed using the Phonopy package.\cite{togoFirstPrinciplesPhonon2015}

\section{Data Availability}
The training data for the universal IAP is available at \url{http://doi.org/10.6084/m9.figshare.19470599}. The phonon dispersion curves of 328 dynamically stable materials are available at \url{http://doi.org/10.6084/m9.figshare.20217212}. All generated hypothetical compounds and their corresponding M3GNet predictions are provided at \url{http://matterverse.ai}.

\section{Code Availability}
The source code for M3GNet is available at \url{https://github.com/materialsvirtuallab/m3gnet}.

\clearpage
\providecommand{\latin}[1]{#1}
\makeatletter
\providecommand{\doi}
  {\begingroup\let\do\@makeother\dospecials
  \catcode`\{=1 \catcode`\}=2 \doi@aux}
\providecommand{\doi@aux}[1]{\endgroup\texttt{#1}}
\makeatother
\providecommand*\mcitethebibliography{\thebibliography}
\csname @ifundefined\endcsname{endmcitethebibliography}
  {\let\endmcitethebibliography\endthebibliography}{}



\section*{Supplementary material}

# Supplementary Information

# A Universal Graph Deep Learning Interatomic Potential for the Periodic Table

Chi Chen* and Shyue Ping Ong*

*Materials Virtual Lab, Department of NanoEngineering, University of California San Diego, 9500 Gilman Dr, Mail Code 0448, La Jolla, CA 92093-0448, United States*

E-mail: chenc273@outlook.com; ongsp@eng.ucsd.edu

# Characteristics of M3GNet models

## Long-range interactions

Different from the MLIAPs, M3GNet can simulate long-range interaction by stacking graph convolution layers without increasing the cutoff radius. To compare the long-range behavior, we used a model system MgO. The ion interactions in the test MgO system are described by the Buckingham potential with long-range electrostatics.

$$V_{ij} = A_{ij} \exp(-r_{ij}/\rho_{ij}) - \frac{C_{ij}}{r_{ij}^6} + \frac{Kq_iq_j}{\epsilon r_{ij}} \qquad (1)$$

where $A_{ij}$, $\rho_{ij}$ and $C_{ij}$ are the Buckingham potential parameters, $q_i$ is the charge for atom $i$, $K$ is the energy-conversion constant and $\epsilon$ is the dielectric constant.

We used the MgO Buckingham potential to simulate the material and generate the long-range force field data. The parameters were take from the work by Shukla et al.[1]

We took the MgO (mp-1265) structure from the Materials Project[2] as the initial structure. Then we applied -0.1, 0.1, and 0.2 isotropic strains to the structures, producing a total of four structures. For each structure, we randomly displace the atoms up to 0.5 Å in each direction for 100 times, producing a total of 400 structures for the MgO data set. The data is split into 80%-20% train-test data. Energies and forces are calculated using the potential described in the previous section using LAMMPS[3] by setting the short-range interaction cutoff as 10 Å and carrying out long-range calculations using the Ewald summation.[4]

M3GNet and MTP models with cutoffs of 5 Å were trained on the train data and evaluated on the test data. We calculated the equation of state (EOS) for MgO using different computational methods, as shown in Figure S1a. The M3GNet in this work outperforms the MTP in extrapolation outside of the training data regimes in general.

The probe the interaction reach, we adopted the locality tests following the ideas proposed by Bartók et al.[5] and reiterated by Deringer and Csányi[6]. For the MgO cubic conventional cell, we produced a 6 × 6 × 6 super cell, with dimensions 25.54 Å × 25.54 Å × 25.54 Å. Then

for the Mg atom at the center, we fix the its neighbor atoms within a radius of $r_{local}$ and introduce random displacements (max 0.2 Å) to the atoms out of $r_{local}$. We recorded the forces on the center atom and compute its standard deviation $\sigma(\boldsymbol{f})$. If $\sigma(\boldsymbol{f}) = 0$ at certain $r_{local}$. The interaction described by the computational method is localized within $r_{local}$.

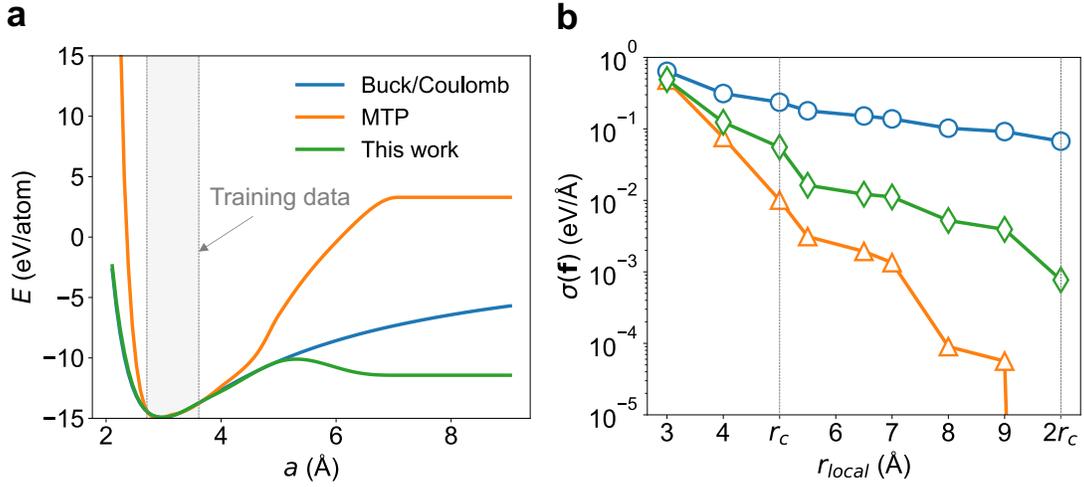

Figure S1: M3GNet compared to MTP in long-range MgO dataset in (a) equation-of-state study and (b) locality tests.

Figure S1b shows the $\sigma(\boldsymbol{f})$ as a function of $r_{local}$ for Buckingham with Coulombic forces, MTP, and M3GNet in this work. As expected, the Buckingham potential with long-range force is highly non-local, with large $\sigma(\boldsymbol{f})$ even at $r_{local} = 10$ Å. The M3GNet in this work shows similar long range behavior and has non-zero $\sigma(\boldsymbol{f})$ at $r_{local} = 10$ Å. On the contrary, the interaction of MTP is localized, with $\sigma(\boldsymbol{f})$ vanishing at $2r_c = 10$ Å. It should be noted that while the MTP interactions in terms of energy is limited to $r_c$, the force impact goes to $2r_c$ because the forces are calculated as follows:

$$\boldsymbol{f}_i = -\frac{E}{\boldsymbol{r}_i} = -\sum_j \frac{E_j}{\boldsymbol{r}_i} \qquad (2)$$

Although all $j$ in Equation 2 is within $r_c$ of atom $i$, for an atom $j$ at the boundary close to $r_c$, the atom $k$ that is $2r_c$ away from $i$ can also contribute to $E_j$. Hence the force $\boldsymbol{f}_i$ will be affected by atom $k$ that is $2r_c$ away. Nevertheless, the force locality for MTP beyond $r_c$

is relatively minimal and vanishes completely at $2r_c$.

In conclusion, the M3GNet model proposed in this work is able to capture long-range interactions well beyond the set cutoff radius.

## Potential smoothness

A critical requirement of an interatomic potential is its prediction smoothness across changes in structures. Figure S2 shows that by applying strains to a Ni cell, the energy, forces, and stresses all change smoothly despite non-smooth changes of the number of bonds defined within a cutoff of 5 Å.

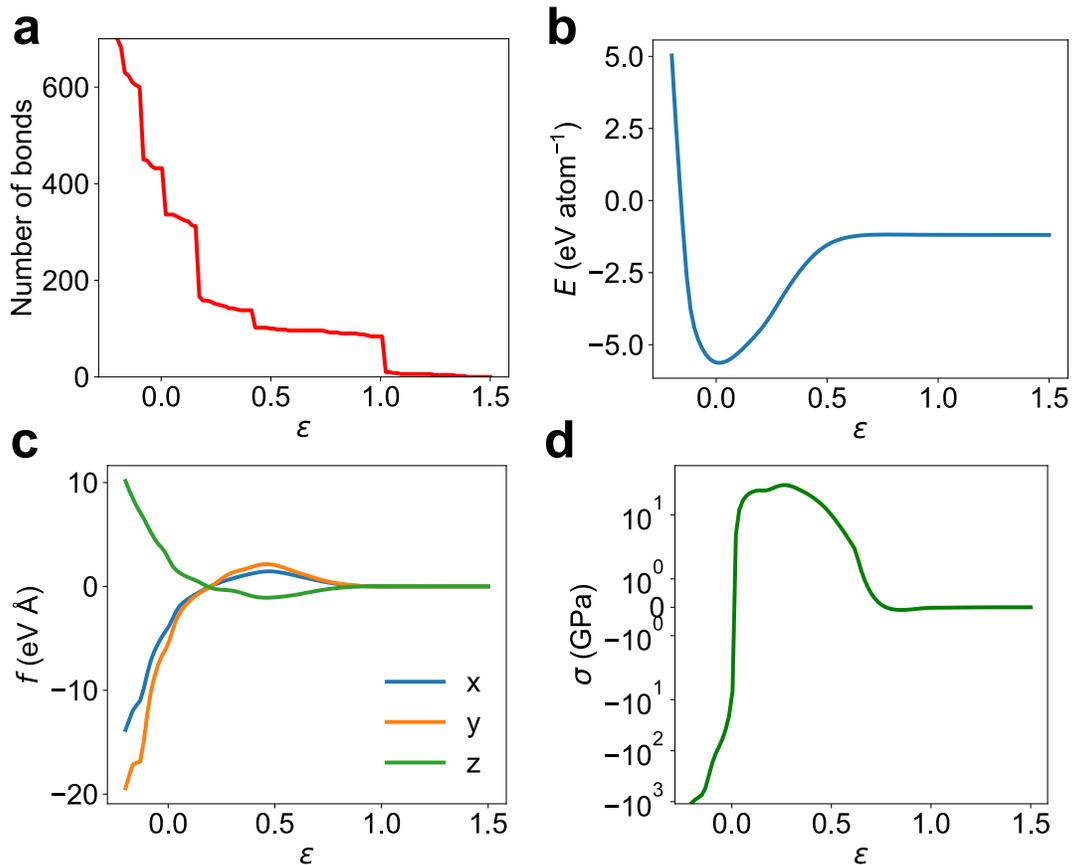

Figure S2: **Continuity of M3GNet models with strain applied on a Ni super cell with atom displacements on the first atom.** **a**, the number of bonds change, **b**, the energy per atom change, **c**, the force on the first atom change and **d**, the hydrostatic stress change. All bonds are broken at strain close to 150%, at which point the energy, force and stress changes continuously.

# MPF.2021.2.8 distribution and M3GNet fitting using different targets

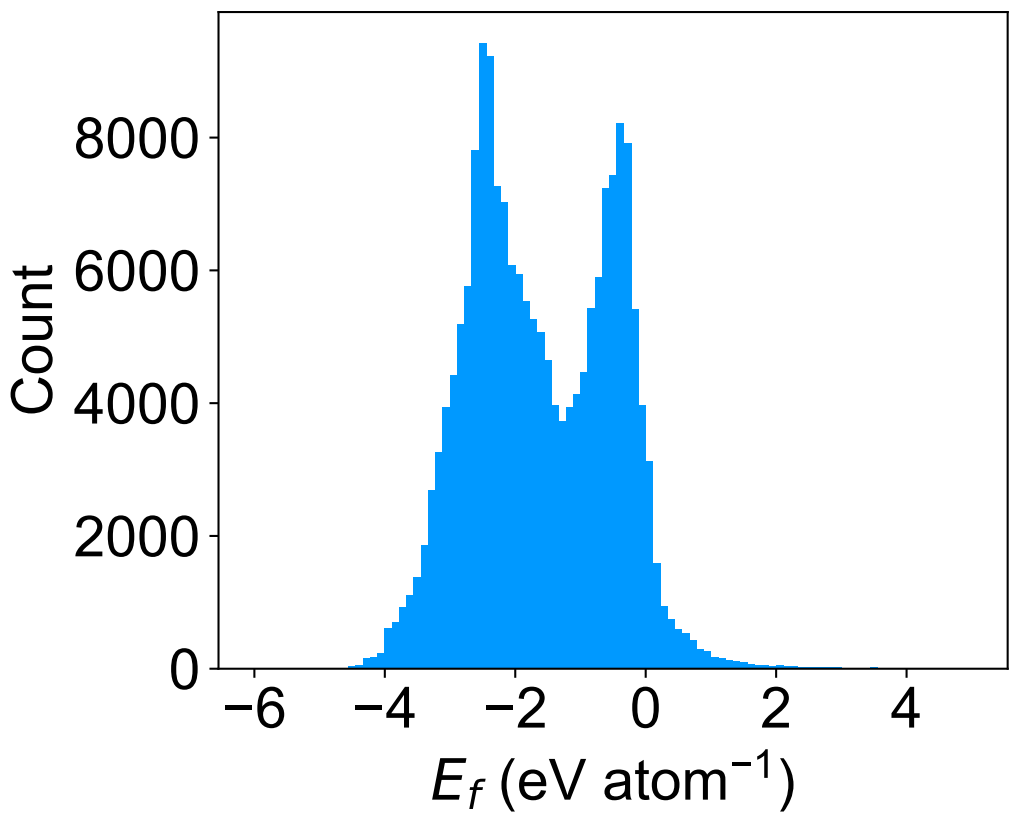

Figure S3: **Formation energy distribution of the MPF.2021.2.8 dataset**

The MPF.2021.2.8 data distribution is shown in Table S1.

Table S1: The mean, standard deviation, min, max and quantiles of the MPF.2021.2.8 dataset.

| Target | Mean | Std | Min | Max | [0.25, 0.5, 0.75] Quantile |
| --- | --- | --- | --- | --- | --- |
| $E$ (eV atom$^{-1}$) | -5.975 | 1.860 | -28.731 | 49.575 | [-7.081, -6.191, -4.938] |
| $f$ (eV Å$^{-1}$) | 0.0 | 3.395 | -2570.567 | 2552.991 | [-0.033, 0.0, 0.033] |
| $\sigma$ (GPa) | -0.833 | 25.13 | -5474.488 | 1397.567 | [-0.049, 0, 0.006] |

Table S2: Mean absolute errors (MAEs) of M3GNet models trained on energy, energy and force, and energy, force, and stress of the MPF.2021.2.8 data. Three models were trained, using energy-only (M3GNet-$E$), energy and force (M3GNet-$EF$), and energy, force and stress (M3GNet-$EFS$). Each error metric is calculated from the mean of three independent model trainings.

| Model | Energy (meV atom$^{-1}$) | Force (meV Å$^{-1}$) | Stress (GPa) |
|---|---|---|---|
| MAD | 1345.6 | 270.3 | 2.068 |
| M3GNet-$E$ | 41.7±2.1 | 379.4±48.7 | 15.05±2.63 |
| M3GNet-$EF$ | 34.0±2.9 | 70.2±1.7 | 0.80±0.12 |
| M3GNet-$EFS$ | 34.7±3.1 | 71.7±1.1 | 0.41±0.01 |

# M3GNet dynamic properties calculations

The phonon dispersion curves and density of state (DOS) calculations for SiO$_2$ polymorphs are shown in Figure S4, where the M3GNet calculated results in Figure S4a are in good agreements with the DFT calculations in Figure S4b, yet the M3GNet calculations took only seconds to run, which are at least four orders of magnitude faster than the DFT calculations. The model calculations used a frozen phonon approach while the original calculations used DFPT. The frozen phonon approach requires larger supercells to produce long wavelength behavior. The supercell size difference also explains the differences in the number of bands. The DOS center versus average atomic mass results in Figure S5 are also in agreement with DFT results.

The bulk moduli and shear moduli calculated from M3GNet in comparison with DFT are shown in Figure S6. The bulk moduli matched well with DFT calculations, but the shear moduli were more challenging to match.

# M3GNet relaxation for materials discovery

The MAEs as a function of M3GNet relaxation steps are shown in Figure S7, where the MAE dropped to 0.047 eV atom$^{-1}$ after about 10 relaxation steps.

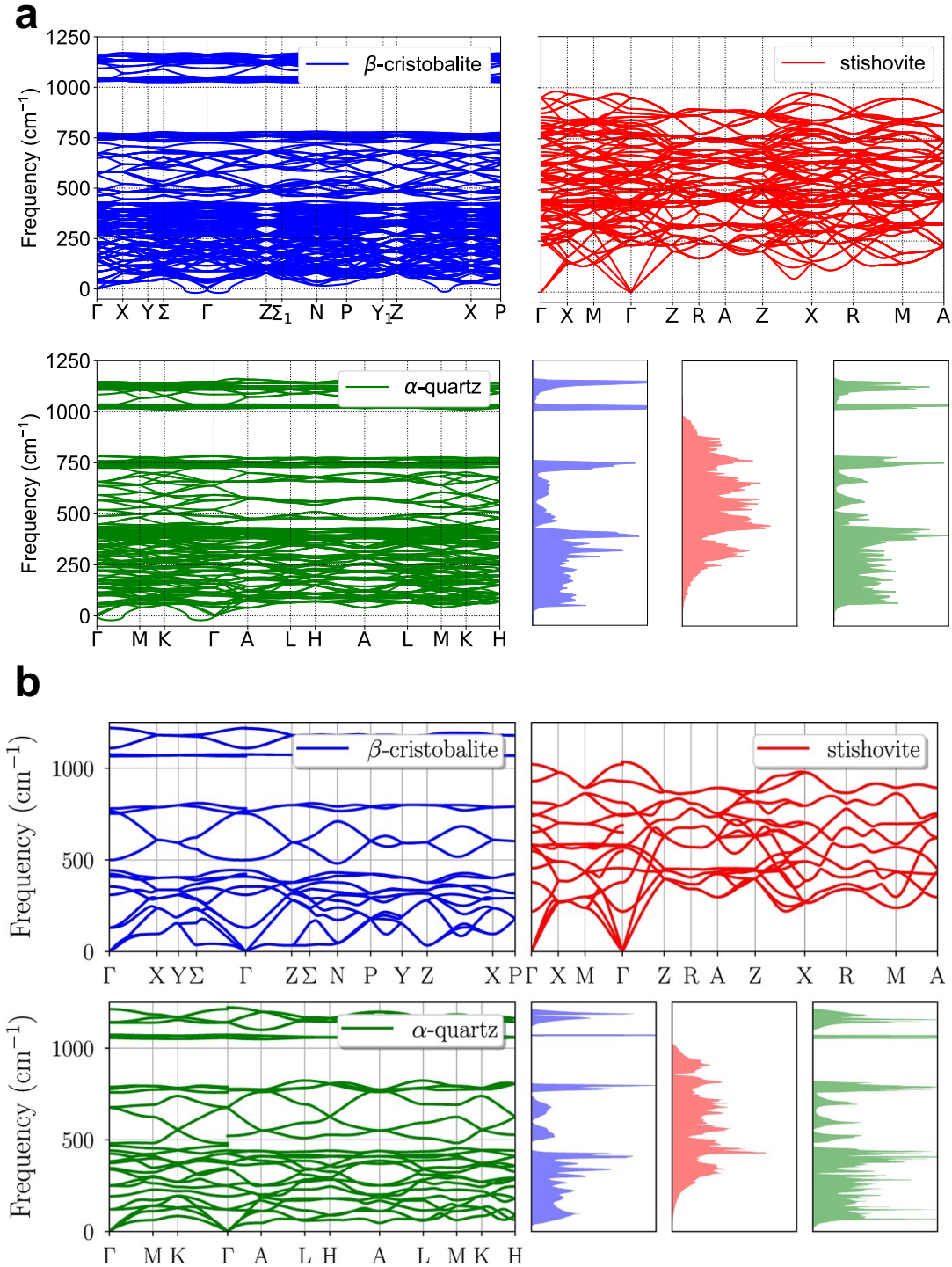

Figure S4: **M3GNet-calculated phonon band structure examples for SiO$_2$ polymorphs**. (a) The band structures for the $\beta$-cristobalite, stishovite, and $\alpha$-quartz structures, and their corresponding density of states (DOS) calculated from M3GNet using 2×2×2 supercells. (b) The same calculations by DFT using the PBEsol functional from Petretto et al.[7] The M3GNet results show quantitative agreements with DFT. Panel (b) is reprinted under the terms of Creative Commons Attribution 4.0 International License. Copyright 2018,

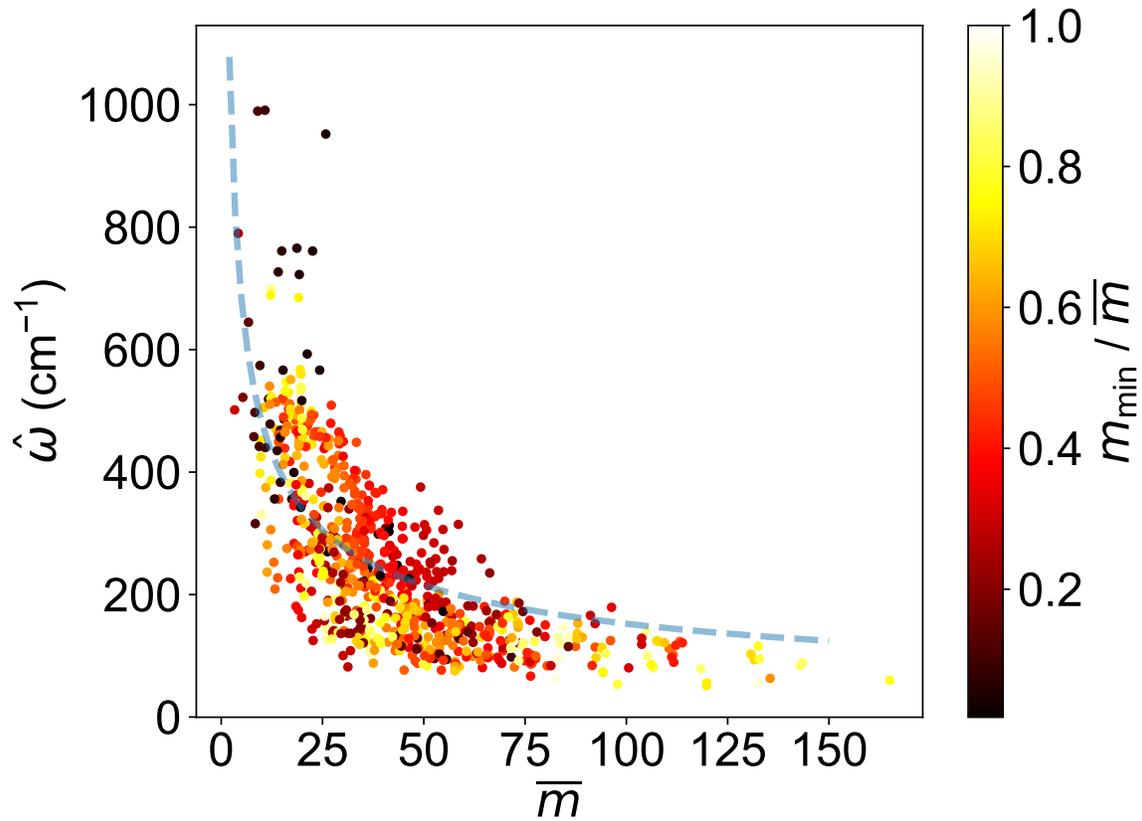

Figure S5: **The predicted phonon DOS center as a function of the average atomic mass** $\overline{m} = \left(\frac{1}{n}\sum_k \sqrt{M_k}\right)^2$, **colored by the ratio between the minimal atomic mass** $m_{\min}$ **and** $\overline{m}$. The fitted line follows the trend $\omega \propto 1/(\overline{m})^2$. The results are in quantitative agreement with the work by Petretto et al.[7]

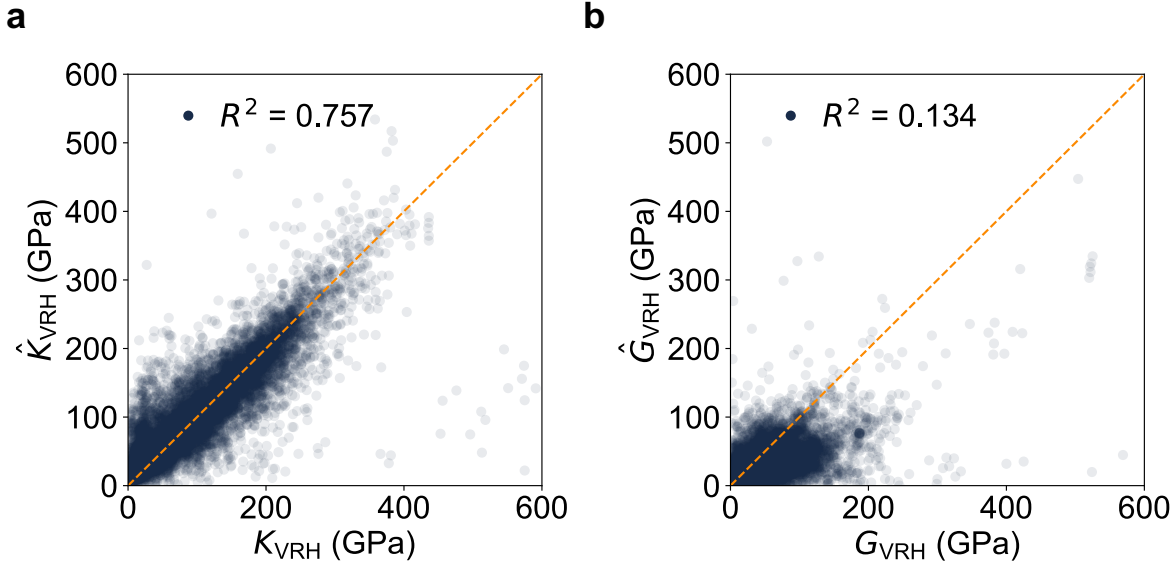

Figure S6: **The bulk and shear moduli calculated by the M3GNet compared to DFT**. The agreement in bulk moduli (a) is better than shear moduli (b).

An example of M3GNet relaxation is shown in Figure S8. The energy, force, and stress all converged within less than 200 steps. At the same time, the lattice constants, angles and the mean atom distances all converged closed to the equilibrium structures from DFT. The XRD pattern of model-relaxed structures also match with that from DFT-relaxation.

Figure S9 shows the M3GNet energy changes during M3GNet relaxation ($\Delta E_{\text{M3GNet}-\text{relax}}$), and DFT energy changes during DFT relaxation ($\Delta E_{\text{DFT}-\text{relax}}$). In both all chemistry category (Figure S9a-b) and oxide category (Figure S9c-d), the DFT energy changes are at least one order of magnitude smaller than the M3GNet energy changes, suggesting that M3GNet effectively relaxes the structures close to their equilibrium.

Figure S10 shows the DFT stable ratio for 1000 structures taken uniformly from the total discovery pool of $\sim 1.8$ million structures based on their $E_{\text{hull}-\text{m}}$ values. The stable ratio (DFT stable structures / structures below $E_{\text{hull}-\text{m}}$ threshold) decreases monotonically with the $E_{\text{hull}-\text{m}}$ threshold, and reaches 0.31 at threshold 0.001 eV atom$^{-1}$.

For materials in Figure S10, we compared the DFT relaxation CPU cost with and without M3GNet pre-relaxation, as shown in Figure S11. On average the DFT relaxation CPU cost without M3GNet pre-relaxation is 2.971 times the one with M3GNet pre-relaxation. In other

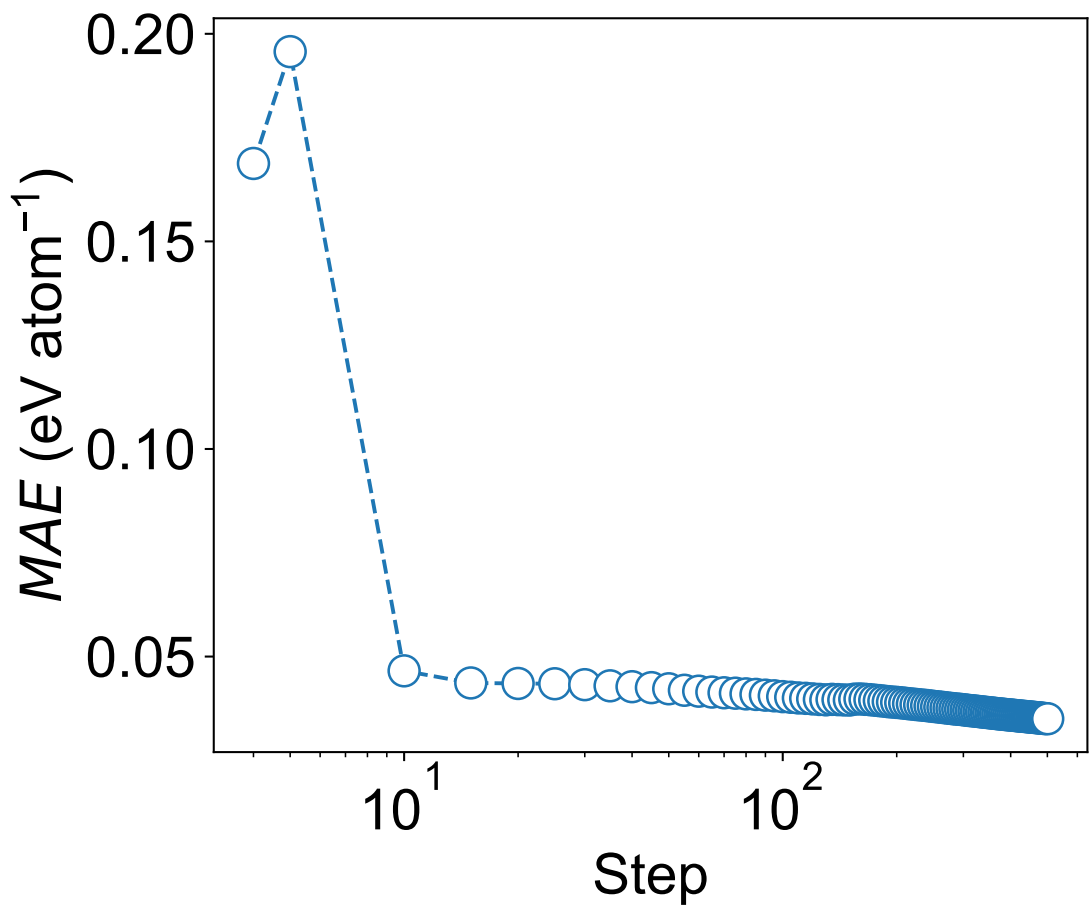

Figure S7: **Relaxation convergence.** The MAE between model calculated energies during relaxation and the final DFT energies as a function of M3GNet relaxation steps.

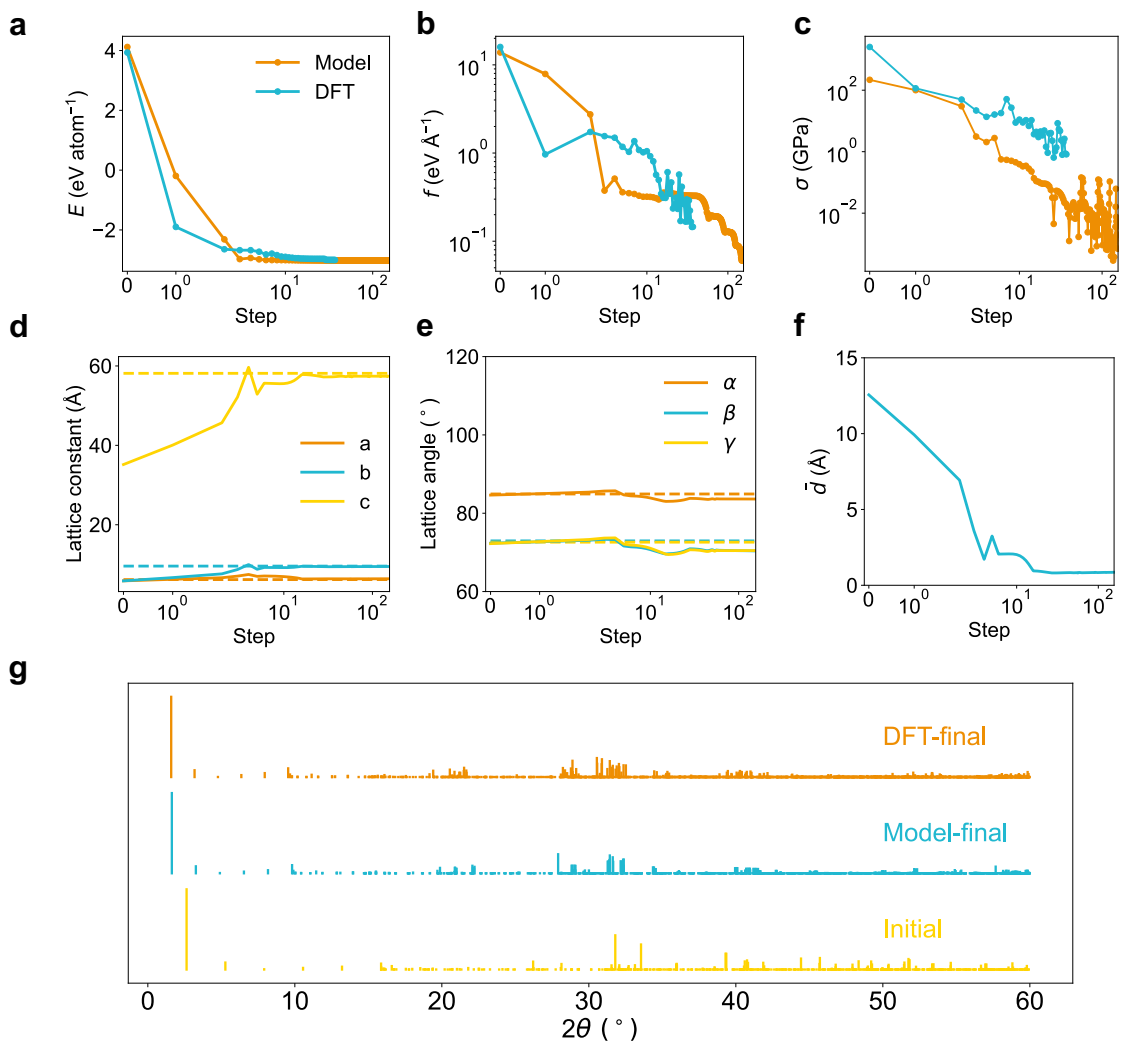

Figure S8: **A structural relaxation example using the model compared to DFT.** **a-c**, The energy, force, and stress evolution during the model optimization, compared to the DFT relaxation for mp-685089 $K_{57}Se_{34}$. **d-e.** The lattice constant and angle changes during relaxation compared to the final DFT structure (dashed lines). **f.** The average atom distances to their final positions. **g.** The X-ray diffraction pattern of the initial structure, and the ones for the DFT relaxed final structure and model relaxed final structure.

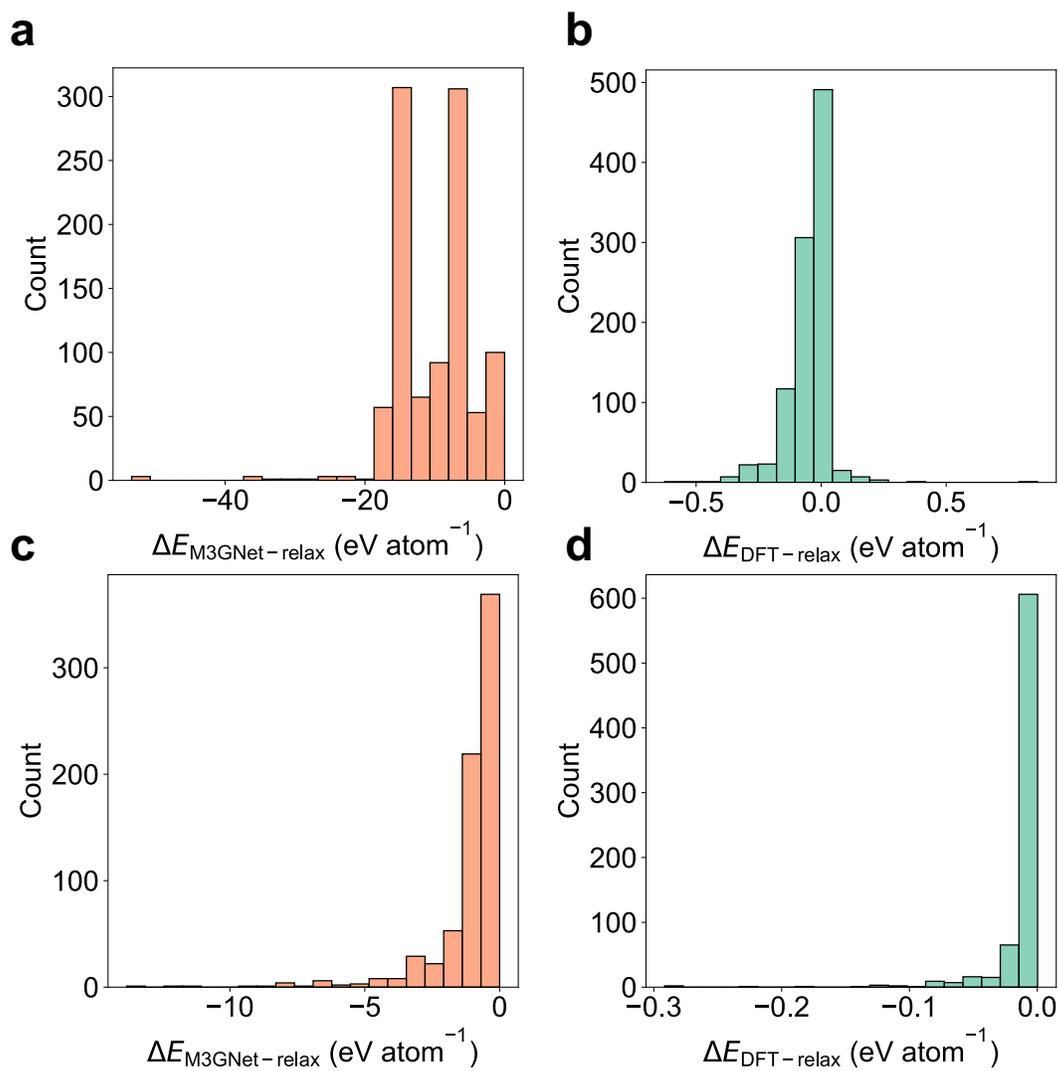

Figure S9: **Distribution of energy changes during M3GNet and DFT relaxations.** **a**, M3GNet energy changes during M3GNet relaxation for materials in All category. **b**, DFT energy changes in subsequent DFT relaxations for materials in All category. **c-d**, same plots as **a-b** but on the Oxide category.

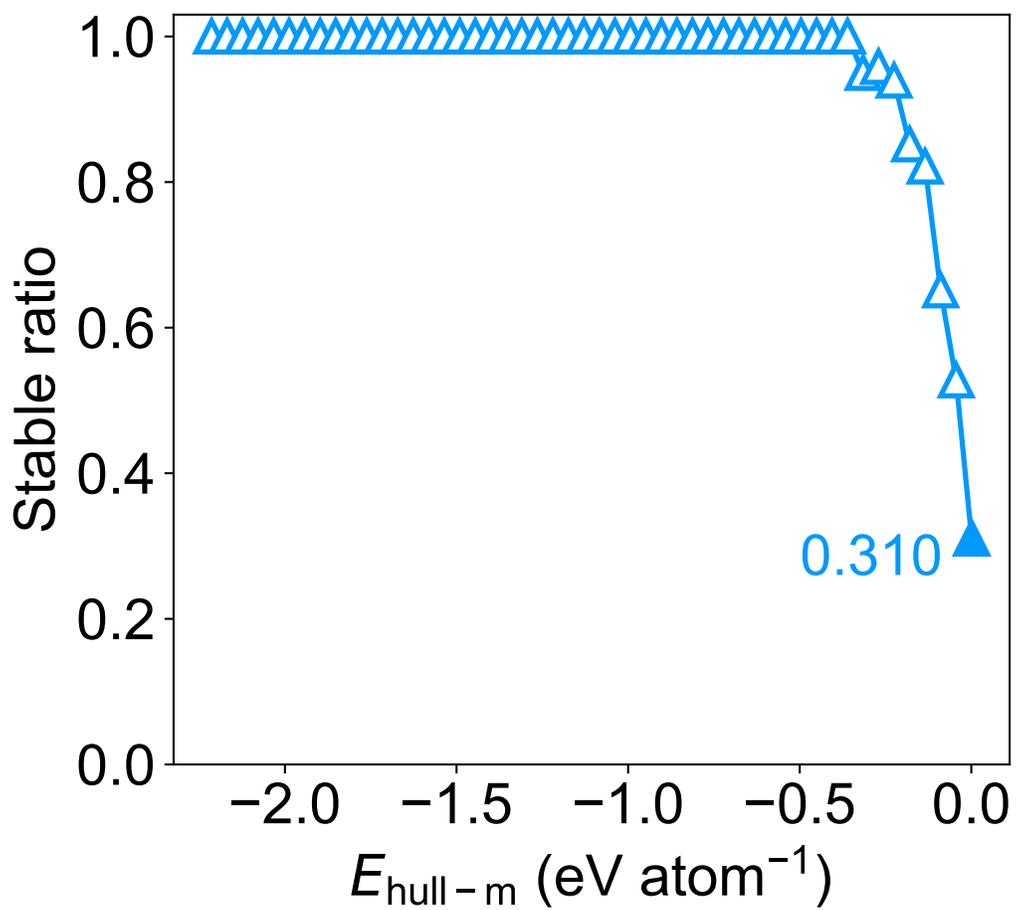

Figure S10: **DFT stable ratio as a function of $E_{\text{hull-m}}$ threshold for a uniform sample of 1000 structures.**

words, the M3GNet relaxation can accelerate DFT relaxation if one aims to obtain all DFT results.

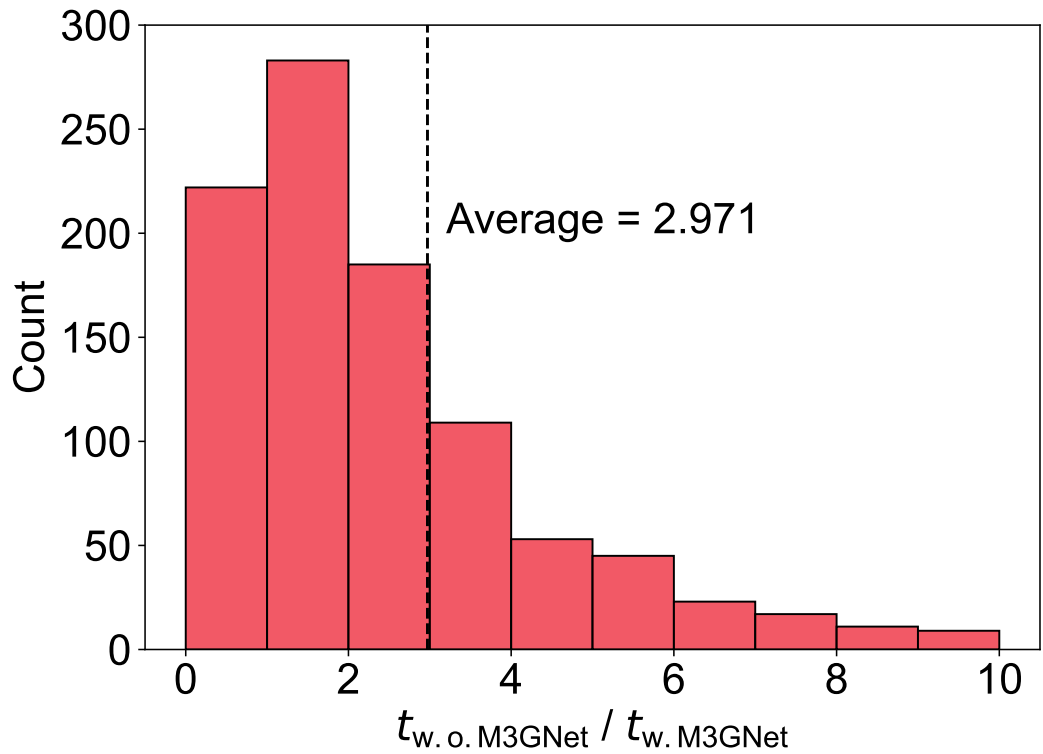

Figure S11: **DFT acceleration with M3GNet pre-relaxation**.

# M3GNet for molecular dynamics simulations

The M3GNet model trained on MPF.2021.2.8 can also be used for molecular dynamics simulations. We chose $Li_3YCl_6$ as an initial study case from our recent work.[8] The M3GNet calculated mean squared displacements (MSDs) for Li, Y, and Cl show reasonable temperature dependencies, as shown in Figure S12. Interestingly, the M3GNet calculated conductivities matched well with AIMD simulations from Wang et al.[9] The extrapolated room-temperature Li conductivity and the activation energy fitted from the Arrhenius equation also showed excellent agreements.

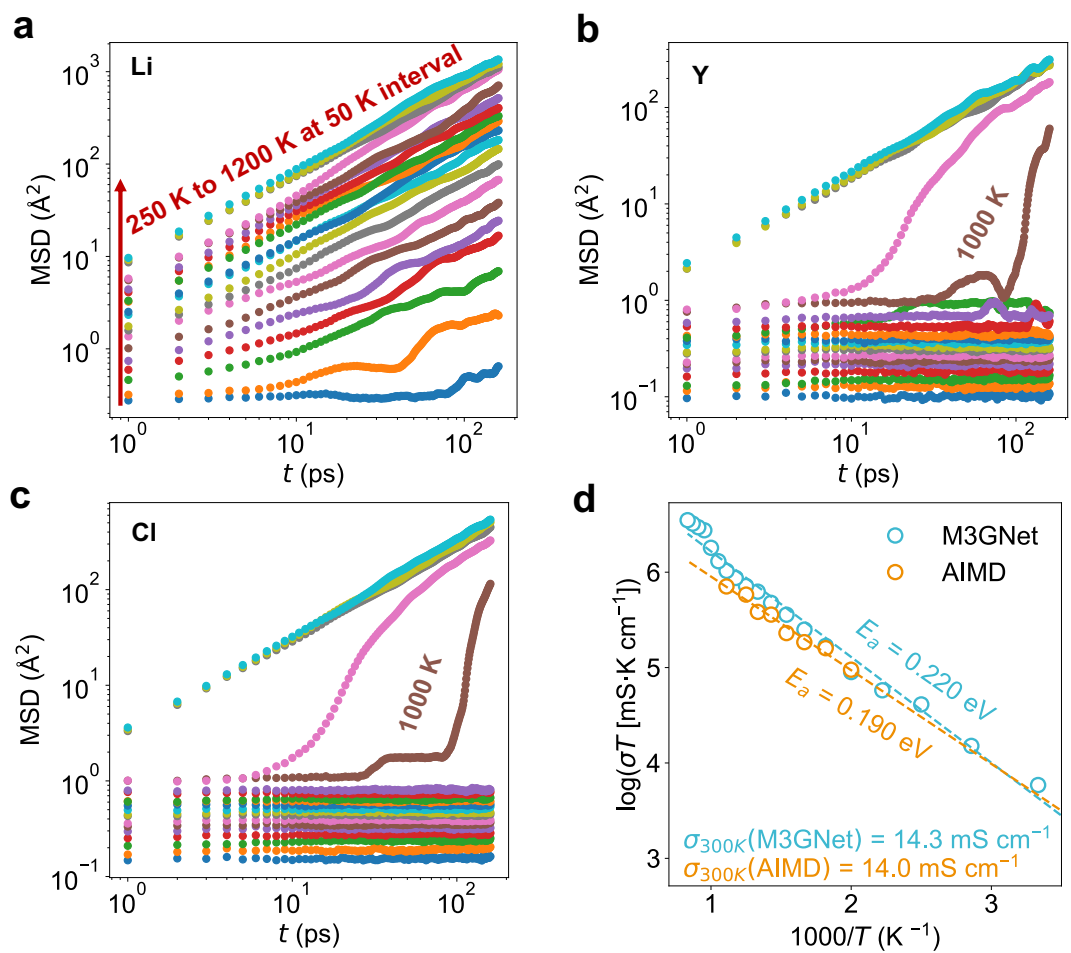

Figure S12: **M3GNet MD results on $Li_3YCl_6$.** **a-c** the mean-squared displacements (MSDs) for Li, Y and Cl, respectively, at temperatures ranging from 250 K to 1200 K at 50 K intervals. The material was found to melt at temperatures between 950 K and 1000 K, as shown by the sudden increase of MSD for framework ions Y and Cl. **d** The Arrhenius plot of the Li conductivity using M3GNet vs AIMD simulations from Wang et al.[9] The M3GNet activation energy of 0.22 eV and extrapolated room temperature conductivity of 14.3 mS cm$^{-1}$ compare well with the AIMD values of 0.19 eV and 14 mS cm$^{-1}$.

To demonstrate M3GNet's utility in screening for unexplored lithium superionic conductors, we performed NVT MD simulations on Li-containing compounds in Materials Project that satisfying the following criteria: 1) energy above hull less than 50 meV atom$^{-1}$ and band gap greater than 2 eV, 2) no transition metals other than Sc, Y, Zr and Nb, and 3) alkaline element fraction greater than 10% in the formula. We also manually added a few entries with known conductivities, such as $Li_3Y(PS_4)_2$, $Li_5P(S_2Cl)_2$, $LiZnPS_4$, $LiAl(PS_3)_2$, studied by Zhu et al.,[10] the tetragonal (t)- and cubic (c)-$Li_7La_3Zr_2O_{12}$, $Na_2B_2H_{12}$, etc. The filtering and additions gave us 837 potential lithium superionic conductors for MD simulations using the short molecular dynamics approach proposed by Zhu et al.,[10] where 25 ps of molecular dynamics are performed on the materials at 800 K and 1200 K temperatures and the MSDs of the diffusive species are recorded.

Figure S13a maps all the Li-containing compounds in our materials pool. The materials on the upper right are desired materials with both high MSDs and low diffusion energy barriers ($\propto \log((MSD - 1200K)/(MSD - 800K))$). The M3GNet model correctly "re-discovered" known Li superionic conductors such as $LiBH_4$, $Li_7P_3S_{11}$, $Li_{10}M(PS_6)_2$ (M=Ge, Si, Sn), $Li_3PS_4$, $Li_3Y(PS_4)_2$, $Li_5P(S_2Cl)_2$, $Li_4GeS_4$, to have high MSDs at both temperatures. The halides such as $Li_3InCl_6$, $Li_3InBr_6$, $Li_3YCl_6$, $Li_3ErCl_6$ are also good conductors but have slightly smaller MSDs than the first batch of superionic conductors. The M3GNet model also predicts that the c-$Li_7La_3Zr_2O_{12}$ has similar MSD with t-$Li_7La_3Zr_2O_{12}$ at 1200 K but much larger MSD at 800 K, suggesting the the cubic phase is a better ionic conductor at low temperature. Oxides such as $Li_4GeO_4$, $Li_4SiO_4$, $Li_2Ge_7O_{15}$, $Li_3ClO$ and $Li_3BrO$ show very small MSDs and hence minimal Li conductivity. However, in experiments, $Li_3ClO$ was known to be a highly Li conductive material. We postulate that defects are the main driving force for high Li conductivity in this material, while in simulation there are no defects. Indeed, high quality deposited thin films with few defects show very small Li conductivity in $Li_3ClO$.[11] AIMD simulations by Zhang et al.[12] also found that pristine $Li_3ClO$ and $Li_3BrO$ are poor conductors, with no Li diffusion up to 1500 K. AIMD that showed Li diffusion was

done with explicit Li vacancies.[13]

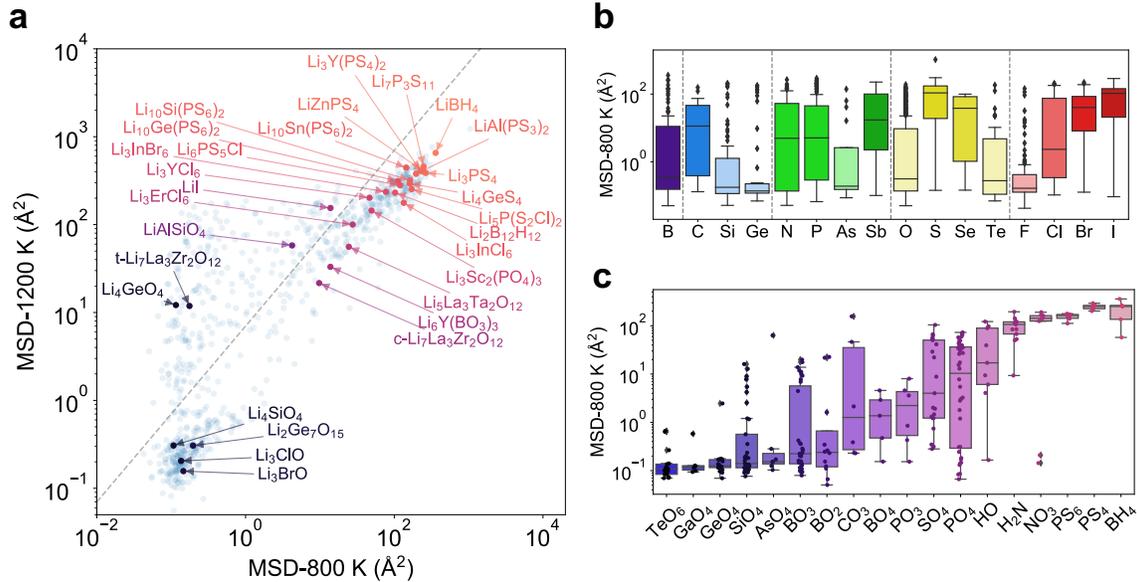

Figure S13: **Mean squared displacements (MSDs) of short-time M3GNet molecular dynamics simulations. a**, The distributions of 800 K and 1200 K MSDs for all Li-containing compounds passing the filtering criteria, supplemented by manual entries. The dashed line has a slope of 7, and materials below this line have desired Li diffusion energy barrier lower than 0.4 eV. **b**, the MSDs at 800 K projected onto the anion types. **c**, the MSDs at 800 K projected onto the anion groups.

Many materials distribute on the top left (low low-temperature MSDs and high Li diffusion energy barriers) and the bottom left (low MSDs at both temperatures). Very few materials distribute on the upper right and below the dashed line with diffusion energy barriers less than 0.4 eV. The results also suggest the difficulties in finding new superionic conductors.

We further group the materials according to their anion group (Figure S13b) and the polyanions (Figure S13c). For carbon group elements (C, Si, and Ge), the Li diffusivities decrease monotonically with the atomic number. For pnictogens, the trend forms a convex shape where arsenide compounds have the lowest overall Li conductivity and antimonides have the highest conductivity. For the chalcogen anions, sulfur stands out in facilitating Li diffusion, followed by selenium. Oxides and tellurides are usually poor Li conductors.

best Li conductors. Comparing the elements in the second row of the periodic table, boron and nitrogen form better lithium conductors than oxygen and fluorine. The projections onto the polyanion groups shown in Figure S13c also suggest that oxides are relatively poor conductors, with the exception of $CO_3$, $SO_4$, $PO_4$ that have several conductive compounds. Not surprisingly, $PS_6$ and $PS_4$ polyanions facilitate fast Li conduction. Lastly, the $BH_4$ polyanion group, which are all $LiBH_4$ polymorphs, show remarkable Li diffusivity. The trends of anions and polyanions match with experimental findings and distill chemical knowledge for guiding new Li superionic conductor design.

## Applications of M3GNet to benchmark materials datasets

### MatBench materials data

We took the nine structural MatBench datasets by Dunn et al.[14] as the material property benchmark. The datasets cover a wide range of materials properties with varying data sizes, including crystal energy, phonon spectrum, optical properties, mechanical properties, etc. We trained the M3GNet models on those properties using the same model architectures as the potential model. The M3GNet results compared to other state-of-the-art models are shown in Table S3. The MODNet[15,16] model shows lower or comparable errors in the small to medium data size regimes. However, for larger data such as the whole formation energy dataset of MP, the original authors acknowledge that the graph models such as MEGNet outperform MODNet by a factor of 2.[16] The M3GNet model, on the other hand, is able to achieve consistently high accuracy across all the datasets, and is more accurate than the previous MEGNet models in all nine tasks. In particular, for the large ($\sim$ 15,000) to very large datasets (>100,000), the M3GNet models achieve much higher accuracy than the rest; the accuracy improvements compared to the previous best graph model are 20.5% for the perovskite formation energy data, 22.1% for the MP band gap, and even 41.3% for the MP formation energy data.

Table S3: Performance of M3GNet models on crystal property predictions compared to other state-of-the-art models using the same test sets as the AutoMatminer work.[14] Mean absolute error (MAE) for regression tasks and area-under-curve (AUC) of receiver operating characteristic curve for classification tasks are used as the model metrics. The lowest errors are highlighted with bold fonts.

| Target, Data Size | M3GNet | MOD-Net[16] | AtomSets-$V_1$[17] | MEG-Net[14,18] | AutoMat-miner[14] |
|---|---|---|---|---|---|
| Regression Tasks | | | | | |
| $E_{exfoliation}$ (meV atom$^{-1}$), 636$^b$ | 50.1±11.9 | **34.5** | **40±5** | 55.9 | **38.6** |
| PhonDOS Peak (cm$^{-1}$), 1265$^c$ | **34.1±4.5** | **38.75** | 54±9 | **36.9** | 50.8 |
| $n$, 4764$^e$ | **0.312±0.063** | **0.297** | 0.46±0.03 | 0.478 | **0.299** |
| log($K_{VRH}$) (GPa), 10987$^f$ | **0.058±0.003** | 0.0548 | 0.07±0.00 | 0.0712 | 0.0679 |
| log($G_{VRH}$) (GPa), 10987$^g$ | 0.086±0.002 | **0.0731** | 0.09±0.00 | 0.0914 | 0.0849 |
| Perovskite $E_f$ (meV atom$^{-1}$), 18928$^h$ | **6.6±0.2** | - | 12±0 | 8.3 | 38.8 |
| MP $E_g$ (eV), 106113$^i$ | **0.183±0.005** | - | 0.25±0.01 | 0.235 | 0.282 |
| MP $E_f$ (meV atom$^{-1}$), 132752$^j$ | **19.5±0.2** | - | 44±1 | 33.2 | 173 |
| Classification Tasks | | | | | |
| MP Metallicity, 106113$^m$ | 0.958±0.001 | - | 0.96±0.00 | **0.977** | 0.909 |

# M3GNet as molecular force fields

The MD17 datasets[19–21] were used as the first test case as M3GNet models as force fields. The MD17 datasets contain several small molecules and their molecular dynamics trajectories with energies and forces computed by density functional theory. The comparisons between M3GNet and other models are shown in Table S4, where for each molecule, only 1000 snapshots were used for training and validation and the rest were used as test for metric reporting. Each error metric in M3GNet is the mean error from three different random sampling and model training.

The M3GNet models achieved consistently lower errors than the less complex EANN[22] models. For sGDML[21] and the deep learning-based models, such as message-passing based DimeNet,[23] and the Equivalent neural networks NequIP[24] and PaiNN,[25] the errors are all low and are comparable to the M3GNet models. No model is able to outperform the rest in

Table S4: Performance of M3GNet models compared to the existing models on the MD17 datasets. In each cell, the errors are reported in mean absolute error (MAE) and are written in energy (meV), force (meV Å$^{-1}$) pairs. The M3GNet results were obtained by average of three independent model fitting using different random seeds for data splitting. The lowest errors are highlighted by bold fonts.

| Molecule | M3GNet | EANN[22] | sGDML[21] | DimeNet[23] | NequIP[24] | PaiNN[25] |
|---|---|---|---|---|---|---|
| Aspirin | 7.8, 18.8 | 14.1, 43.0 | 8.4, 29.4 | 8.8, 21.6 | -, 15.1 | **7.2, 14.7** |
| Benzene[19] | 3.5, **8.1** | - | **3.1**, 8.8 | 3.4, **8.1** | -, **8.1** | - |
| Benzene[21] | **4.0, 1.9** | - | 4.2, 2.4 | - | -, 2.3 | - |
| Ethanol | **2.4, 5.3** | 4.4, 20.3 | 3.1, 14.5 | 2.8, 10.0 | -, 9.0 | 2.8, 9.7 |
| Malonaldehyde | **3.4, 9.5** | 5.9, 26.9 | 4.3, 18.0 | 4.5, 16.6 | -, 14.6 | 4.3, 14.9 |
| Naphthalene | **5.1**, 5.4 | 6.1, 22.3 | **5.0**, 4.9 | 5.3, 9.3 | -, 10.3 | **5.0, 3.3** |
| Salicylic acid | **5.1**, 10.1 | 6.1, 22.3 | **5.0**, 12.2 | 5.8, 16.2 | -, **4.2** | **5.0**, 8.5 |
| Toluene | **4.0**, 4.7 | 4.8, 16.6 | **4.2**, 6.2 | 4.4, 9.4 | -, 4.4 | 4.1, **4.1** |
| Uracil | **4.5**, 7.0 | 4.8, 15.3 | **4.7**, 10.5 | 5.0, 13.1 | -, 7.5 | **4.6, 6.0** |

all molecules.

**MD17 CCSD/CCSD(T) benchmark**

The quantum-accurate CCSD/CCSD(T)[19,21] calculations of the molecules in the MD17 datasets are used for further molecular force field benchmark. The M3GNet model results are compared to the available sGDML,[21] PhysNet[26] and NequIP,[24] as shown in Table S5. In this dataset, the M3GNet model achieves excellent accuracies across all the the molecules, and has state-of-the-art accuracy in Ethanol, Malonaldehyde and Paracetamol. M3GNet also shows state-of-the-art accuracies in energies of Aspirin and Toluene, and forces of Azobenzene and Benzene.

Table S5: Performance of M3GNet models compared to the existing models on the MD17 CCSD/CCSD(T) datasets.[19,21] In each cell, the errors are reported in mean absolute error (MAE) and are written in energy (meV), force (meV Å$^{-1}$) pairs. The M3GNet results were obtained by average of three independent model fitting using different random seeds for data splitting. The lowest errors are highlighted by bold fonts.

| Molecule | M3GNet | sGDML[21] | PhysNet[26] | NequIP[24] |
|---|---|---|---|---|
| Aspirin | **6.0**, 19.7 | 6.8, 33.0 | - | -, **14.7** |
| Azobenzene | 7.5, **12.2** | **4.0**, 17.7 | 8.5, 20.1 | - |
| Benzene | 0.15, **1.06** | **0.13**, 1.8 | - | -, 0.8 |
| Ethanol | **0.92, 5.51** | 2.2, 15.0 | - | -, 9.4 |
| Malonaldehyde | **1.3, 8.2** | 1.6, 10.7 | - | -, 16.0 |
| Paracetamol | **6.6, 14.5** | 6.6, 21.3 | 7.9, 22.5 | - |
| Toluene | **1.0**, 5.5 | 1.3, 8.9 | - | -, **4.4** |



\end{document}